\documentclass{article}

\usepackage{amsmath}
\usepackage{amssymb}
\usepackage{graphicx}
\usepackage{bm}
\usepackage{mathdots}
\usepackage{booktabs}
\usepackage{rotating}
\usepackage[ruled,linesnumbered]{algorithm2e}
\usepackage[a4paper,left=2.7cm,right=2.7cm,top=2.5cm,bottom=2.5cm]{geometry}
\usepackage{fancyhdr}
\usepackage[framemethod=tikz]{mdframed}
\newmdtheoremenv[%
backgroundcolor=green!10,%
outerlinecolor=black,%
leftmargin=0,%
rightmargin=0,
innertopmargin =3pt,%
innerleftmargin = 5pt,
innerrightmargin = 5pt,
splittopskip = \topskip,%
skipabove = \baselineskip,%
skipbelow = \baselineskip,%
roundcorner=5, ntheorem]
{theorem}{Theorem}[section]
\newtheorem{corollary}{Corollary}[section]
\newtheorem{proposition}{Proposition}[section]
\newtheorem{lemma}{Lemma}[section]
\newtheorem{definition}{Definition}[section]
\makeatletter 
\@addtoreset{equation}{section}
\makeatother  

\newtheorem{remark}{Remark}[section]

\begin{document}
\title{{\LARGE \bf Enhanced total variation minimization for stable image reconstruction}}

\author{Congpei An\footnotemark[1]
       \quad\quad Hao-Ning Wu\footnotemark[2]
       \quad\quad  Xiaoming Yuan\footnotemark[2]}

\renewcommand{\thefootnote}{\fnsymbol{footnote}}
\footnotetext[1]{School of Mathematics, Southwestern University of Finance and Economics, Chengdu, China (ancp@swufe.edu.cn).}
\footnotetext[2]{Department of Mathematics, The University of Hong Kong, Hong Kong, China (\{hnwu,xmyuan\}@hku.hk).}


\maketitle

\begin{abstract}
The total variation (TV) regularization has phenomenally boosted various variational models for image processing tasks. We propose to combine the backward diffusion process in the earlier literature of image enhancement with the TV regularization, and show that the resulting enhanced TV minimization model is particularly effective for reducing the loss of contrast. The main purpose of this paper is to establish stable reconstruction guarantees for the enhanced TV model from noisy subsampled measurements with two sampling strategies, non-adaptive sampling for general linear measurements and variable-density sampling for Fourier measurements. In particular, under some weaker restricted isometry property conditions, the enhanced TV minimization model is shown to have tighter reconstruction error bounds than various TV-based models for the scenario where the level of noise is significant and the amount of measurements is limited. Advantages of the enhanced TV model are also numerically validated by preliminary experiments on the reconstruction of some synthetic, natural, and medical images.
\end{abstract}

\textbf{Keywords: }{total variation, image reconstruction, backward diffusion, anisotropic, loss of contrast, stability, difference-of-convex regularization}

\medskip

\textbf{AMS subject classifications.}   94A08, 94A20, 68U10, 68Q25

\bigskip

\section{Introduction}
Since the work of Rudin, Osher and Fatemi \cite{rudin1992nonlinear}, various variational models based on the total variation (TV) have been intensively studied for image processing problems; see, e.g., \cite{MR2731599,chambolle2016introduction} for reviews. Given linear measurements $y\in\mathbb{C}^m$ observed via
\begin{equation}\label{equ:measurements}
y={\mathcal{M}} {\bar{X}} +e
\end{equation}
from an unknown image ${\bar{X}}\in\mathbb{C}^{N\times N}$, where ${\mathcal{M}}:\mathbb{C}^{N\times N}\rightarrow\mathbb{C}^m$ is a linear operator defined component-wisely by
\begin{equation*}
[{\mathcal{M}}({\bar{X}})]_j:=\langle M_j,{\bar{X}}\rangle=\text{tr}(M_j{\bar{X}}^*),
\end{equation*}
for suitable matrices $M_j$ with $m$ considerably smaller than $N^2$, and $e\in\mathbb{C}^m$ is a noise term bounded by $\|e\|_2\leq \tau$ with level $\tau\geq0$, reconstruction of the unknown $\bar{X}$ can be modeled as the following TV minimization problem:
\begin{equation}\label{equ:TVmodel}
              \min_{X\in\mathbb{C}^{N\times N}}~\|X\|_{\text{TV}}\quad\text{s.t.}\quad \|{\mathcal{M}} X-y\|_2\leq \tau,
\end{equation}
where $\|\cdot\|_{\text{TV}}$ is the TV semi-norm. Note that the TV semi-norm can be mainly categorized as the isotropic \cite{chambolle2004algorithm} and anisotropic \cite{chambolle2005total} cases for discrete images. In this paper, we discuss how to enhance the canonical constrained TV model (\ref{equ:TVmodel}) by the recently proposed springback regularization in \cite{an2021springback} for image reconstruction, and establish stable reconstruction guarantees.

As profoundly analyzed in \cite{needell2013stable}, the constrained TV model \eqref{equ:TVmodel} has the advantage of reconstructing high-quality images from a relatively small number of measurements. Theoretical analysis in \cite{needell2013stable} is mainly based on the seminal compressed sensing (CS) works \cite{candes2006robust,donoho2006compressed}. Note that the classic CS theory assumes the sparsity of the (vector) signal of interest or its coefficients under certain transformation, and correspondingly the signal reconstruction can be modeled as some $\ell_1$-norm minimization problems. The CS theory can be extended to image reconstruction because natural images usually have (approximately) sparse gradients. Indeed, mathematically the TV semi-norm of a discrete image $X\in\mathbb{C}^{N\times N}$ is just the sum of the magnitudes of its gradient $\nabla X\in\mathbb{C}^{N\times N \times 2}$. That is,
\begin{equation}\label{equ:TV}
\|X\|_{\text{TV}_a}:=\|\nabla X\|_1,
\end{equation}
where the definition of $\nabla X$ can be found in Section \ref{sec:basics}. We note here that the definition \eqref{equ:TV} leads to the anisotropic version of the TV semi-norm. Since the anisotropic and isotropic TV semi-norms are equivalent up to a factor of $\sqrt{2}$ (see an explanation in Section \ref{sec:basics}), as \cite{needell2013stable}, we only consider the anisotropic case for succinctness and the following discussion can be extended to the isotropic case analogously.

Models using the $\ell_1$-norm are fundamental to various CS problems, while solutions to such models may be over-penalized because the $\ell_1$ regularization tends to underestimate high-amplitude components of the solution, as analyzed in \cite{fan2001variable}. Accordingly, many non-convex alternatives have been proposed in the literature to overcome this pitfall and thus promote sparsity more firmly; see, e.g., the $\ell_p$ ($0<p<1$) regularization \cite{chartrand2007exact,foucart2009sparsest}, the $\ell_{1-2}$ regularization \cite{yin2015minimization}, and the transformed $\ell_1$ regularization \cite{zhang2018minimization}. The non-convexity feature in image processing has also been emphasized in various papers, see, e.g., \cite{MR3560068}. Recently, we proposed the springback regularization in \cite{an2021springback}, and it can be generalized as the following for discrete images:
       \begin{equation}\label{equ:penalty}
              \mathcal{R}_{\alpha}(X):=\|\nabla X\|_1-\frac{\alpha}{2}\|\nabla X\|_2^2,
       \end{equation}
where $\alpha>0$ is a meticulously-chosen parameter to ensure the positiveness or the well-definedness of \eqref{equ:penalty}, and $\|\nabla X\|_2^2$ is the sum of the squared magnitudes of $\nabla X$. Note that the springback regularization \eqref{equ:penalty} is of difference-of-convex. To some extent, it keeps both the nice recoverability of various non-convex surrogates of the TV regularization and the computability of the original TV regularization. To be consistent with the TV literature, we call \eqref{equ:model} an \emph{enhanced TV} regularization in this paper.

Non-convex penalties proposed in the CS literature are mainly rooted in the field of statistics, and they are usually applied in straightforward ways in the image processing literature. Interestingly, as elaborated in Section \ref{sec:first}, the enhanced TV regularization \eqref{equ:penalty} has some intrinsic interpretations from the perspective of image processing. We are thus encouraged to consider the enhanced TV model 
       \begin{equation}\label{equ:model}
              \min_{X\in\mathbb{C}^{N\times N}}~\mathcal{R}_{\alpha}(X)\quad\text{s.t.}\quad \|{\mathcal{M}} X-y\|_2\leq \tau
       \end{equation}
for image reconstruction, and we aim at establishing some stable reconstruction guarantees theoretically. It is worth noting that, despite the theoretical reconstruction guarantees established in \cite{an2021springback} for sparse signals or signals that are sparse after an orthonormal transform, the guarantees established in \cite{an2021springback} are not applicable to the enhanced TV model \eqref{equ:model}. The reason is that the gradient transform $\nabla:X\rightarrow \nabla X$ fails to be orthonormal, as mentioned in \cite{needell2013stable}. Also, we notice that the idea of enhancing the TV regularization (the isotropic version) with a subtraction of a squared norm of the image gradient was skated over in \cite{mollenhoff2015primal}, and it was empirically tested for some image denoising problems despite the lack of rigorous study for reconstruction guarantees from a few measurements.

\subsection{An image processing view of the enhanced TV regularization}\label{sec:first}

Solutions to TV-based models may lose contrast across edges. That is, the contrast of the regions on both sides of an edge may be reduced, and thus blur may occur near the edge. We refer the reader to \cite{benning2013higher,strong2003edge} for discussions on the loss of contrast caused by various image processing models using TV regularization.

Partial differential equations (PDEs) and variational approaches have been intensively investigated to enhance the contrast. On the PDE side, some well-known approaches were proposed to tackle the loss of contrast for image enhancement. For example, the shock filter was proposed in \cite{osher1990feature} to deal with blur-like image degradations, creating strong discontinuities at image edges and flattening the image within homogeneous regions. Afterwards, the shock filter has been generalized in many ways, see, e.g., \cite{alvarez1994signal,welk2007theoretical}. Another important example is the forward-and-backward (FAB) diffusion scheme proposed in \cite{gilboa2002forward} to simultaneously remove the noise and enhance the contrast. Since then, a number of influential works regarding the FAB diffusion have been conducted, see, e.g., \cite{welk2009theoretical,welk2005pde,welk2018discrete}. Despite that different PDE schemes were designed, a common feature of these works is that the \emph{backward diffusion process} is adopted to enhance the contrast of the edges in a concerning image. Since backward diffusion is a classical example of an ill-posed problem \cite{MR0455365}, most of these PDE schemes sound numerically challenging; we refer the reader to \cite{chambolle2021approximating,chambolle2021learning,welk2008locally} on how to discretize and solve these PDEs efficiently. On the variational side, it was shown in \cite{chan2005aspects,nikolova2002minimizers,MR3560068} that the contrast of the edges could be enhanced by using non-smooth data fidelity terms, which can be achieved by, e.g., replacing the squared $\ell_2$-norm data fidelity term with the $\ell_1$-norm. There are also some attempts to add negative terms into the variational model to maximize the contrast, see, e.g., \cite{galdran2015enhanced,pierre2017variational}, though their connections with the TV regularization are not considered.

\begin{figure}[htbp]
  \centering
  \includegraphics[width=14cm]{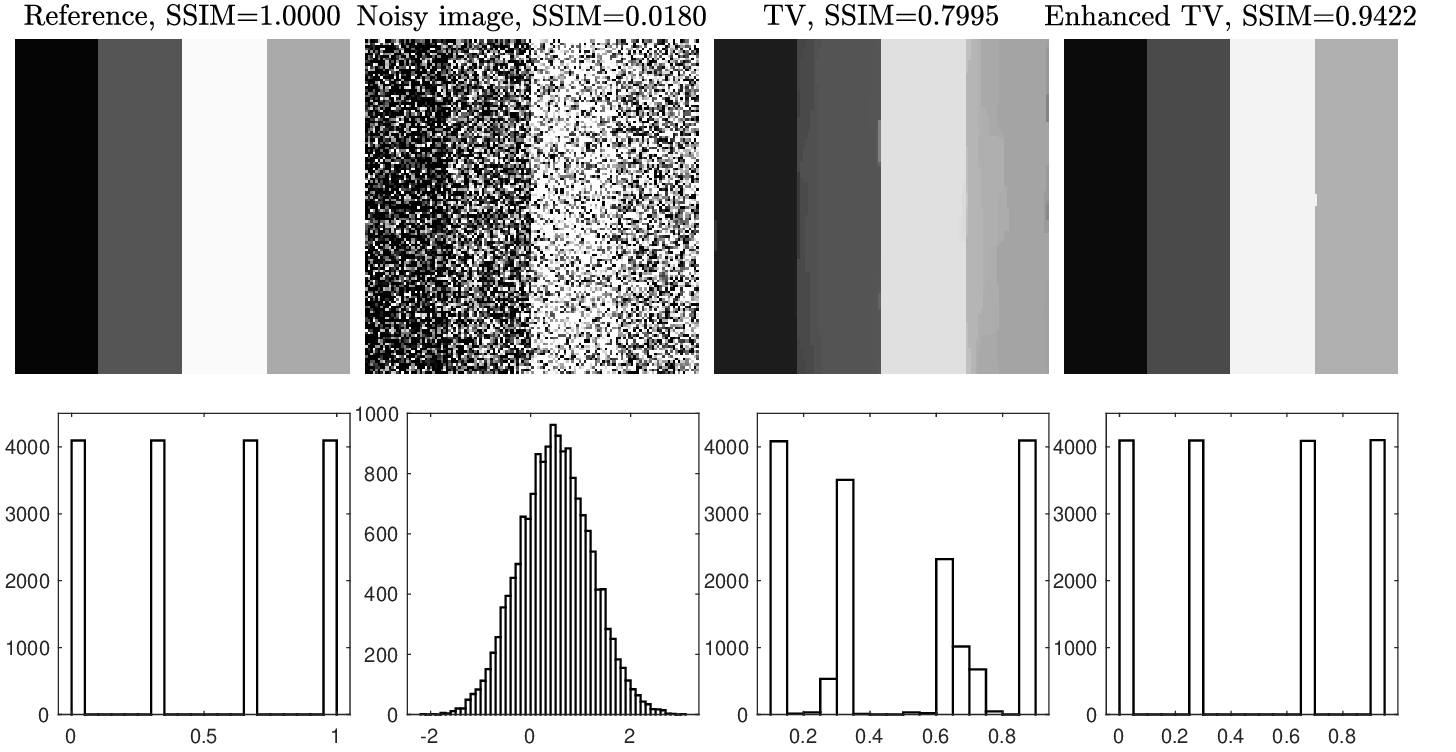}
  \caption{Illustration of the TV and enhanced TV regularization for image denoising. First row: SSIM values of each image; Second row: histograms of pixel intensities of each image.}\label{fig:motivation}
\end{figure}

We remark that the enhanced TV model \eqref{equ:model} is related to the backward diffusion from the PDE perspective. An explanation in the context of the Euler--Lagrange (E--L) equation in a continuum setting is included in Appendix \ref{sec:continuum}. Briefly speaking, the term $-\frac{\alpha}{2}\|\nabla X\|_2^2$ generates an additional backward diffusion term $-\alpha\Delta X$ into the E--L equation corresponding to the classic TV regularization. In Figure \ref{fig:motivation}, we empirically illustrate that the enhanced TV regularization \eqref{equ:penalty} is very effective for some fundamental denoising and deblurring problems. Figure \ref{fig:motivation} clearly shows that the enhanced TV regularization \eqref{equ:penalty} outperforms the original TV regularization in removing noise, reducing loss of contract, and maintaining the smoothness inside homogeneous regions. These convincing performances are clear motivations for us to consider theoretical reconstruction guarantees for the enhanced TV model \eqref{equ:model}. Implementation details for reproducing Figure \ref{fig:motivation} are enclosed in Appendix \ref{sec:implementation}.

In Figure \ref{fig:motivation}, we also note that the enhanced TV regularization \eqref{equ:penalty} may not perfectly overcome another drawback of TV: the \emph{staircase effect}. That is, solutions to TV-based models may have stair-like edges. There are many efforts trying to avoid this effect, including the replacement of the TV regularization with an exponentiation term of it \cite{blomgren1997total}, the usage of the infimal convolution of functionals with first and second order derivatives as regularizer \cite{chambolle1997image}, the addition of some higher-order terms into the E--L equation corresponding to the variational TV model \cite{chan2000high}, the total generalized variation \cite{bredies2010total}, the usage of some modified infimal convolutions \cite{MR2544023,setzer2011infimal} regarding \cite{chambolle1997image}, and many others.

In a nutshell, the enhanced TV regularization \eqref{equ:penalty} can be interpreted as introducing an additional backward diffusion term into the underlying E--L equation corresponding to the TV regularization for the purpose of enhancing the contrasts along the edges.

\subsection{Contributions}\label{sec:contributions}
In the CS context, it is possible to \emph{exactly} recover a signal if the signal is exactly sparse and its measurements are noise-free; otherwise, we can only establish \emph{stable} recovery guarantees. The term \emph{stable} in this paper is mainly concerned with both inexact sparsity and measurement noise. Our analysis is conducted under the \emph{restricted isometry property} (RIP) framework studied in \cite{candes2005decoding}. We say that a linear operator $\mathcal{A}:\mathbb{C}^{n_1\times n_2}\rightarrow\mathbb{C}^m$ has the RIP of order $s$ and level $\delta\in(0,1)$ if
\begin{equation}\label{equ:RIP}
(1-\delta)\|X\|_2^2\leq\|\mathcal{A} X\|_2^2\leq(1+\delta)\|X\|_2^2\quad\forall s\text{-sparse }X\in\mathbb{C}^{n_1\times n_2},
\end{equation}
and the smallest $\delta$ for \eqref{equ:RIP} is said to be the \emph{restricted isometry constant} (RIC) associated with $\mathcal{A}$.

We first investigate \emph{non-adaptive} subsampled linear RIP measurements of an image ${\bar{X}}\in\mathbb{C}^{N\times N}$ with noise level $\tau>0$. By ``non-adaptive,'' we mean that the sampling strategy is not designed with specific structures or under certain distributions. In Theorem \ref{thm:1}, we show that the enhanced TV model \eqref{equ:model} can stably reconstruct an image ${\bar{X}}\in\mathbb{C}^{N\times N}$ from some non-adaptive subsampled linear RIP measurements which are contaminated by noise, with the RIP order $\mathcal{O}(s)$, the RIP level $\delta<0.6$, and the noise level $\tau>0$. Moreover, the required RIP level $\delta<1/3$ derived in \cite{needell2013stable} for the TV model \eqref{equ:TVmodel} is weakened to $\delta<0.6$ for the enhanced TV model \eqref{equ:model} under the additional condition \eqref{equ:thm1alpha} for the parameter $\alpha$. We also show in Theorem \ref{thm:2} that the reconstruction error bound in Theorem \ref{thm:1} can be further improved if more measurements are allowed.

The above reconstruction guarantees for non-adaptive measurements require the subsampled measurements and the Haar wavelet basis to be sufficiently incoherent. This requirement is satisfied by many kinds of measurements except for the Fourier frequency measurements, because low-order wavelets and Fourier measurements are highly correlated, as analyzed in \cite{krahmer2013stable}. Fourier measurements play essential roles in many imaging tasks. For example, as discussed in \cite{fannjiang2010compressed,krahmer2013stable}, the measurement process of various image processing procedures such as radar, sonar, and computer tomography can be modeled (with appropriate approximation and discretization) by taking samples from weighted discrete Fourier transforms. It is also known (see, e.g., \cite{lustig2008compressed}) that measurements taken for magnetic resonance imaging (MRI) can be well modeled as Fourier coefficients of the desired image.

On the other hand, many empirical pieces of evidence, including the first works \cite{lustig2007sparse,lustig2008compressed} for compressed sensing MRI, have shown that better reconstruction quality is possible by subsampling Fourier frequency measurements with a preference for low frequencies over high frequencies. Thus, we follow the \emph{density-variable} sampling strategy proposed in \cite{krahmer2013stable} and choose Fourier measurements randomly according to an \emph{inverse square law density}. We show that from at least $m\gtrsim s\log^3(s)\log^5(N)$ such subsampled Fourier measurements with $s\gtrsim \log(N)$, the enhanced TV model \eqref{equ:model} reconstructs an unknown image ${\bar{X}}$ stably with high probabilities. We also show that the least amount of Fourier measurements required by the enhanced TV model \eqref{equ:model} is only $\left(0.6/(1/3)\right)^{-2}\approx 30.86\%$ of that by the TV model \eqref{equ:TVmodel} as established in \cite{krahmer2013stable}.

\subsection{Related works}
We briefly review some TV-related works on image reconstruction. The reconstruction of a one-dimensional image in $\mathbb{C}^N$ with an exactly $s$-sparse gradient from noise-free, uniformly subsampled Fourier measurements were considered in \cite{candes2006robust}, without stability analysis concerning the inexact sparsity or noise. It was shown that this one-dimensional image could be recovered exactly by solving the corresponding TV model with high probabilities, provided that the number of measurements $m$ satisfies $m\gtrsim s\log(N)$. The reconstruction of a one-dimensional image using noisy measurements was then considered in \cite{cai2015guarantees}. Stability of the reconstruction of approximately sparse images from noisy measurements was first shown in \cite{needell2013stable} for two-dimensional images and soon extended to higher-dimensional cases in \cite{needell2013near}. More specifically, it was asserted in \cite{needell2013stable} that, from some non-adaptive subsampled linear RIP measurements of an image ${\bar{X}}\in\mathbb{C}^{N\times N}$ with the RIP order $\mathcal{O}(s)$, the RIP level $\delta<1/3$, and the noise level $\tau>0$, the solution ${X^{\text{opt}}}$ to the TV model \eqref{equ:TVmodel} satisfies
\begin{equation}\label{equ:TVrecovery}
\|\bar{X}-X^{\rm{opt}}\|_2 \lesssim \log\left(\frac{N^2}{s}\right)\left(\frac{\|\nabla {\bar{X}} -(\nabla {\bar{X}})_s\|_1}{\sqrt{s}}+\tau\right),
\end{equation}
where $(\nabla {\bar{X}})_s$ is the best $s$-sparse approximation to the discrete gradient $\nabla X$. Moreover, with more measurements, it was shown in \cite{needell2013stable} that the log factor in the bound \eqref{equ:TVrecovery} could be removed, and thus the bound \eqref{equ:TVrecovery} can be improved as
\begin{equation}\label{equ:TVrecoveryimproved}
\|\bar{X}-X^{\rm{opt}}\|_2 \lesssim\frac{\|\nabla {\bar{X}} -(\nabla {\bar{X}})_s\|_1}{\sqrt{s}}+\tau.
\end{equation}

In comparison with the bound \eqref{equ:TVrecoveryimproved}, the reconstruction error bound for the enhanced TV model \eqref{equ:model} in Theorem \ref{thm:2} is tighter if the level of noise $\tau$ is relatively large and the number of measurements $m$ is limited. More discussions can be found in Section \ref{sec:discussion}. Besides, the RIP level is assumed to satisfy $\delta<1/3$ in \cite{needell2013stable} for the TV model \eqref{equ:TVmodel}, while we weaken it to $\delta<0.6$ for the enhanced TV model \eqref{equ:model}. Though $\delta < 1/3$ can be improved, as remarked in  \cite{needell2013stable}, the reconstruction error bounds \eqref{equ:TVrecovery} and \eqref{equ:TVrecoveryimproved} for the TV model \eqref{equ:TVmodel} tend to be infinity if $\delta\rightarrow0.6$ (cf. the proof of Proposition 3 in \cite{needell2013stable}). On the other hand, the bounds in Theorems \ref{thm:1} and \ref{thm:2} for the enhanced TV model \eqref{equ:model} are still reasonably valid when $\delta\rightarrow 0.6$; meanwhile, the upper bound required for $\alpha$ tends to be $0$ correspondingly. Thus, as $\delta\rightarrow0.6$, the bounds \eqref{equ:bound1} and \eqref{equ:bound2} in Theorems \ref{thm:1} and \ref{thm:2} for the enhanced TV model \eqref{equ:model} assert the stability of the TV model \eqref{equ:TVmodel} in image reconstruction from a few linear RIP measurements.

As mentioned, guarantees for non-adaptive measurements require the subsampled measurements and the Haar wavelet basis to be sufficiently incoherent. Thus, the mentioned guarantees in \cite{needell2013near,needell2013stable} cannot be directly applied to the situation of Fourier measurements. The first results on image reconstruction from Fourier measurements were derived in \cite{krahmer2013stable} and \cite{poon2015role}, in which \emph{uniform} and \emph{non-uniform}\footnote{In the context of compressed sensing, a \emph{uniform reconstruction guarantee} indicates that a single random draw of a given measurement operator suffices to recover all sparse or approximately sparse vectors. In contrast, a \emph{non-uniform recovery guarantee} states that a single random draw is sufficient for recovery of a fixed vector.} reconstruction guarantees are considered, respectively. More specifically, the approach in \cite{krahmer2013stable} requires a larger number of measurements than \cite{poon2015role}, while its reconstruction error bound is sharper than that in \cite{poon2015role}. In \cite{krahmer2013stable}, uniform reconstruction guarantees were derived for two-dimensional images from noisy Fourier measurements, chosen randomly according to an inverse square law density. Specifically, from at least $m\gtrsim s\log^3(s)\log^5(N)$ such subsampled Fourier measurements with $s\gtrsim \log(N)$, the reconstruction error bound for the TV model \eqref{equ:TVmodel} was derived in the same form of \eqref{equ:TVrecoveryimproved}. We refer to, e.g., \cite{adcock2021improved,adcock2017breaking,krahmer2017total}, for more discussions. As we focus on the uniform reconstruction from non-adaptive measurements, we follow the approach in \cite{krahmer2013stable} to consider Fourier measurements.

\subsection{Outline}
The rest of this paper is organized as follows. In the next section, we summarize some notation and technical backgrounds. In Section \ref{sec:theory}, we establish stable image reconstruction guarantees for the enhanced TV model \eqref{equ:model} from non-adaptive subsampled linear RIP measurements and variable-density subsampled Fourier measurements, respectively. Proofs of the results in Section \ref{sec:theory} are presented in Section \ref{sec:guarantee}. In Section \ref{sec:numerical}, we report some numerical results when the enhanced TV model \eqref{equ:model} is applied to some image reconstruction problems. Different kinds of images with subsampled Fourier measurements are tested. Finally, we make some conclusions in Section \ref{sec:conclusion}.

\section{Preliminaries}

We first summarize some notation and recall some preliminary technical backgrounds.

\subsection{Notation}
For any $x,y\in\mathbb{R}^n$, let $\left<x,y\right>=x^{\text{T}}y$ be their inner product, and $\|x\|_p$ ($p\geq 1$) be the usual $\ell_p$-norm of $x$. For a matrix $X\in\mathbb{R}^{m\times n}$, let $\text{supp}(X):=\{(j,k):X_{j,k}\neq 0\}$ be the support of $X$, and $\|X\|_0$ be the cardinality of $\text{supp}(X)$. $X$ is said to be $s$-sparse if $\|X\|_0\leq s$. Let
\begin{equation*}
\|X\|_{p,q}:=\left(\sum_{j=1}^m\left(\sum_{k=1}^n|X_{j,k}|^p\right)^{q/p}\right)^{1/q}
\end{equation*}
be the entry-wise $\ell_{p,q}$ norm ($p,q\geq1$) of $X$. If $p=q$, $\|X\|_{p,p}$ is denoted by $\|X\|_p$ for short. In particular, the $\ell_{2,2}$ norm is also known as the Frobenius norm, which is induced by the inner product $\langle X,Y\rangle :=\sum_{j=1}^m\sum_{k=1}^n X_{i,j}Y_{i,j}=\text{tr}(XY^*)$ for any $X,Y\in\mathbb{C}^{m\times n}$, where $X^*$ denotes the adjoint of the matrix $X$. For an index set $S\subset\{1,2,\ldots,m\}\times \{1,2,\ldots,n\}$, let $X_{S}\in\mathbb{R}^{m\times n}$ be the matrix with the same entries as $X$ on indices $S$ and zero entries on indices $S^c$. The only exception is ${\mathcal{F}}_{\Omega}$. We denote by ${\mathcal{F}}_{\Omega}$ the restriction of the bivariate discrete Fourier transform ${\mathcal{F}}$ to a subset $\Omega\subset\{-N/2+1,\ldots,N/2\}^2$. Logarithm without indicating base is with respect to base 2. For matrices or vectors $x$ and $y$ of the same dimension, $x\circ y$ denotes the Hadamard (entry-wise) product between $x$ and $y$. We use the notation $a \lesssim  b$ to mean that there exists $C > 0$ such that $a \leq  Cb$, and likewise for the symbol $\gtrsim$.

\subsection{Gradients and TV semi-norms}\label{sec:basics}
For any image $X\in\mathbb{C}^{N\times N}$ represented by an $N\times N$ block of pixel intensities with all intensities $X_{j,k}$ in $[0,1]$, the discrete directional derivatives of $X\in\mathbb{C}^{N\times N}$ are defined in a pixel-wise manner as
\begin{equation*}\begin{split}
 &X_x:\mathbb{C}^{N\times N}\rightarrow\mathbb{C}^{(N-1)\times N}, \quad (X_x)_{j,k}:=X_{j+1,k}-X_{j,k},  \\
 &X_y:\mathbb{C}^{N\times N}\rightarrow\mathbb{C}^{N\times (N-1)}, \quad (X_y)_{j,k}:=X_{j,k+1}-X_{j,k}.
 \end{split}\end{equation*}
The discrete gradient transform $\nabla:\mathbb{C}^{N\times N}\rightarrow \mathbb{C}^{N\times N\times 2}$ is defined in a matrix form as
\begin{equation*}
[\nabla X]_{j,k}:=
\begin{cases}
\left((X_x)_{j,k},(X_y)_{j,k}\right),&1\leq j\leq N-1,~1\leq k\leq N-1,\\
\left(0,(X_y)_{j,k}\right),&1\leq j=N,~1\leq k\leq N-1,\\
\left((X_x)_{j,k},0\right),&1\leq j\leq N-1,~k=N,\\
\left(0,0\right),&j=k=N.
\end{cases}
\end{equation*}
Recall the definition \eqref{equ:TV} of the TV semi-norm. This definition of $\nabla X$ leads to the \emph{anisotropic TV semi-norm}, defined in \cite{chambolle2005total}, that is, the sum of the magnitudes of its discrete gradient
\begin{equation*}
\|X\|_{\text{TV}_a} :=\sum\nolimits_{j,k}\left|[\nabla X]_{j,k}\right|=\sum\nolimits_{j,k}|(X_x)_{j,k}|+|(X_y)_{j,k}|.
\end{equation*}
If the choice of $\left((X_x)_{j,k},(X_y)_{j,k}\right)$ in the definition of $\nabla X$ is replaced by $(X_x)_{j,k}+i(X_y)_{j,k}$, then it leads to the \emph{isotropic TV semi-norm} as defined in \cite{chambolle2004algorithm}:
\begin{equation*}
\|X\|_{\text{TV}_i} :=\sum\nolimits_{j,k}\left((X_x)_{j,k}^2+(X_y)^2_{j,k}\right)^{1/2}.
\end{equation*}

If we regard $\nabla X$ as an $N^2\times 2$ matrix, then $\|X\|_{\text{TV}_a} $ and $\|X\|_{\text{TV}_i} $ are the $\ell_{1,1}$ and $\ell_{2,1}$ norms of $\nabla X$, respectively. Note that both TV semi-norms are equivalent subject to a factor of $\sqrt{2}$:
\begin{equation}\label{equ:uptoafactor}
\|X\|_{\text{TV}_i} \leq\|X\|_{\text{TV}_a} \leq \sqrt{2}\|X\|_{\text{TV}_i} .
\end{equation}
 Moreover, note that $\|\nabla X\|_2=(\sum\nolimits_{j,k}(X_x)_{j,k}^2+(X_y)^2_{j,k})^{1/2}$ in the second component of the enhanced TV regularization \eqref{equ:penalty} is the $\ell_{2,2}$ norm of $\nabla X$.

\subsection{Haar wavelet system}
The Haar wavelet system provides a simple yet powerful sparse approximation of digital images. The following descriptions on this system can be found in, e.g., \cite{needell2013stable}. The \emph{univariate} Haar wavelet system is a complete orthonormal system of square-integrable functions on the unit interval, consisting of the constant function
\begin{equation*}
H^0(t)=\begin{cases}
1,&0\leq t<1,\\
0,&\text{otherwise},
\end{cases}
\end{equation*}
the mother wavelet
\begin{equation*}
H^1(t)=\begin{cases}
1,&0\leq t<1/2,\\
-1,&1/2\leq t<1,
\end{cases}
\end{equation*}
and the dyadic dilations and translates of the mother wavelet $H_{j,k}(t)=2^{j/2}H^1(2^jt-k)$, $j\in\mathbb{N}$, $0\leq k<2^j$. The \emph{bivariate} Haar wavelet system is an orthonormal system for the space $L_2(Q)$ of square-integrable functions on the unit square $Q=[0,1)^2$, and it is derived from the univariate Haar system by tensor product. The bivariate Haar system consists of the constant function and all functions
\begin{equation*}
H_{j,k}^{\ell}(u,v)=2^jH^{\ell}(2^jx-k),\quad j\geq 0,~k\in\mathbb{Z}^2\cap2^jQ,~\ell\in V:=\left\{\{0,1\},\{1,0\},\{1,1\}\right\},
\end{equation*}
where
$H^{\ell}(u,v)=H^{\ell_1}(u)H^{\ell_2}(v)$ and $\ell=(\ell_1,\ell_2)\in V$. Discrete images are isometric to the space $\Sigma_N\subset L_2(Q)$ of piecewise-constant functions
\begin{equation}\label{equ:isometric}
\Sigma_N=\left\{f\in L_2(Q):\quad f(u,v)=c_{j,k},\quad\frac{j-1}{N}\leq u<\frac{j}{N},\quad\frac{k-1}{N}\leq v<\frac{k}{N}\right\}
\end{equation}
with $c_{j,k}=NX_{j,k}$. If $N=2^n$, then the bivariate Haar basis is restricted to the $2^n\times2^n=N^2$ basis functions $\{H_{j,k}^{\ell}:~j\leq n-1\}$ and identified as some discrete images $h_{j,k}^{\ell}$ via \eqref{equ:isometric} forms an orthonormal basis for $\mathbb{C}^{N\times N}$. For any given $\ell=(\ell_1,\ell_2)\in V$, we denote by ${\mathcal{H}}$ the bivariate Haar transform $X\mapsto(\langle X,h_{j,k}^{\ell}\rangle)_{j,k}$. By a slight abuse of notation, we also denote by ${\mathcal{H}}$ the unitary matrix representing this bivariate Haar transform. That is, we denote by ${\mathcal{H}} X$ the matrix product that generates $(\langle X,h_{j,k}^{\ell}\rangle)_{j,k}$.

Some properties of the bivariate Haar wavelet system are summarized below, and the proofs can be found in \cite{needell2013stable}.
\begin{lemma}\label{lem:1}
Suppose $X\in\mathbb{C}^{N\times N}$ is mean-zero, and let $c_{(k)}(X)$ be the bivariate Haar coefficient of $X$ having the $k$th largest magnitude, or the entry of the bivariate Haar transform ${\mathcal{H}} X$ having the $k$th largest magnitude. Then, for all $k\geq 1$, $|c_{(k)}(X)|\leq \tilde{C}\|\nabla X\|_{1}/k$,
where $\tilde{C}>0$ is some constant.
\end{lemma}
\begin{lemma}\label{lem:2}
Let $N=2^n$. For any indices $(j,k)$ and $(j,k+1)$, there are at most $6n$ bivariate Haar wavelets which are not constant on these indices, i.e., $|h^{\ell}_{j,k}(j,k+1)-h^{\ell}_{j,k}(j,k)|>0$.
\end{lemma}
\begin{lemma}\label{lem:3}
The bivariate Haar wavelets satisfy $\|\nabla h^{\ell}_{j,k}\|_1\leq 8$ for all $j,k,\ell$.
\end{lemma}

\subsection{Discrete Fourier system}\label{sec:fouriersystem}

In addition to general RIP measurements, we particularly investigate Fourier measurements. Let $N=2^n$ be a power of 2, where $n\in\mathbb{N}^+$. The following facts of Fourier basis and transform in the context of imaging can be found in, e.g., \cite{krahmer2013stable}. The \emph{univariate} discrete Fourier basis of $\mathbb{C}^N$ consists of vectors
\begin{equation*}
\varphi_k(t)=\frac{1}{\sqrt{N}}e^{i2\pi tk/N},\quad -N/2+1\leq t\leq N/2,
\end{equation*}
indexed by the discrete frequencies in the range of $ -N/2+1\leq k\leq N/2$. The \emph{bivariate} discrete Fourier basis of $\mathbb{C}^{N\times N}$ is a tensor product of univariate bases, i.e.,
\begin{equation*}
\varphi_{j,k}(u,v)=\frac{1}{N}e^{i2\pi (ju+kv)/N},\quad -N/2+1\leq u,v\leq N/2,
\end{equation*}
indexed by the discrete frequencies in the range of $ -N/2+1\leq j,k\leq N/2$.

We denote by ${\mathcal{F}}$ the bivariate discrete Fourier transform $X\mapsto(\langle X,\varphi_{k_1,k_2}\rangle)_{k_1,k_2}$. Again, by a slight abuse of notation, we denote by ${\mathcal{F}}$ the unitary matrix representing this linear map. That is, we denote by ${\mathcal{F}} X$ the matrix product that generates $(\langle X,\varphi_{k_1,k_2}\rangle)_{k_1,k_2}$. Moreover, since limited measurements are considered, we denote by ${\mathcal{F}}_{\Omega}$ the restriction of $F$ to a subset of frequencies $\Omega\subset\{-N/2+1,\ldots,N/2\}^2$.

\section{Main results}\label{sec:theory}
We now establish reconstruction guarantees for the enhanced TV model \eqref{equ:model} from non-adaptive linear RIP measurements and variable-density Fourier measurements, respectively. The following proposition generalizes Theorem 4.1 in \cite{an2021springback} for signal recovery, and it allows us to bound the norm of an image $D$ when it is close to the null space of an RIP operator.

\begin{proposition}\label{prop:main}
Let $\gamma\geq1$, $\delta<0.6$, $\beta_1>0$, $\beta_2>0$, and $\varepsilon\geq0$. Suppose that $\mathcal{A}$ has the RIP of order $k+4k\gamma^2$ and level $\delta$, and that the image $D\in\mathbb{C}^{N\times N}$ satisfies the tube constraint
\begin{equation}\label{equ:tube}
\|\mathcal{A} D\|_2\leq\varepsilon.
\end{equation}
Suppose further that for a subset $S$ of cardinality $|S|\leq k$, $D$ satisfies the cone constraint
\begin{equation}\label{equ:cone}
\|D_{{S}^c}\|_1\leq \gamma\|D_{S}\|_1-\frac{\beta_1}{2}\|D\|_2^2+\sigma+\beta_2\left<E_1,E_2\right>,
\end{equation}
where $E_1,E_2$ could be scalars, vectors, or matrices, and $E_2$ is assumed to satisfy $\|E_2\|_2=\|D\|_2$. Here $\|\cdot\|_2$ denotes the absolute value for scalars, the usual $\ell_2$ vector norm for vectors, and the $\ell_{2,2}$ norm (Frobenius norm) for matrices. If $\beta_2$ satisfies the \emph{posterior verification}
\begin{equation}\label{equ:alpha}
\beta_2\leq \frac{\gamma\sqrt{k}}{2K_2\|E_1\|_2},
\end{equation}
then it holds that
\begin{equation}\label{equ:D2}
\|D\|_2\leq \sqrt{\frac{\gamma\sqrt{k}K_1}{\beta_1 K_2}\varepsilon+\frac{2}{\beta_1}\sigma}\lesssim \sqrt{\frac{\gamma\sqrt{k}}{\beta_1}\varepsilon+\frac{1}{\beta_1}\sigma},
\end{equation}
where
$$K_1:=\frac{3}{2\sqrt{1-\delta}-\sqrt{1+\delta}} \quad\text{and}\quad K_2:=\frac{\sqrt{1+\delta}}{4}\left(K_1+\frac{1}{\sqrt{1+\delta}}\right).$$
Furthermore, we have
\begin{equation}\label{equ:D1D2}
\|D\|_1\leq\frac{(2K_2+1)\gamma\sqrt{k}+2K_2\sqrt{k}}{2K_2}\sqrt{\frac{\gamma\sqrt{k}K_1}{\beta_1 K_2}\varepsilon+\frac{2}{\beta_1}\sigma}+\sigma\lesssim \gamma\sqrt{k}\sqrt{\frac{\gamma\sqrt{k}}{\beta_1 }\varepsilon+\frac{1}{\beta_1}\sigma}+\sigma.
\end{equation}
\end{proposition}
\begin{corollary}\label{rem:main}
There is a linear term of $\sigma$ in \eqref{equ:D1D2}. If $\|D\|_2\geq \sqrt{2\sigma/\beta_1}$, which is compatible with \eqref{equ:D2}, then this linear term can be removed. This corollary will be proved after Proposition \ref{prop:main}.
\end{corollary}

\begin{remark}\label{rem:reasonable}
In the proof of Proposition \ref{prop:main}, we need to ensure $\sqrt{1-\delta}-\sqrt{1+\delta}/2>0$, and this is where the requirement $\delta<0.6$ for the RIP level stems from. Since
\begin{equation}\label{equ:10}
\lim_{\delta\rightarrow 0.6}\frac{K_1}{K_2}=\lim_{\delta\rightarrow 0.6}\frac{4}{\sqrt{1+\delta}+1/K_1}=10,
\end{equation}
the bounds on $\|D\|_2$ and $\|D\|_1$ are still reasonable as $\delta\rightarrow0.6$. As the whole analysis below rests upon Proposition \ref{prop:main}, this fact \eqref{equ:10} suggests that the following reconstruction error bounds \eqref{equ:bound1}, \eqref{equ:bound2}, and \eqref{equ:bound3} are all reasonable as $\delta\rightarrow0.6$.

\end{remark}

\begin{remark}
If $\mathcal{A}$ is assumed to have the RIP of order $5k\gamma^2\geq k+4k\gamma^2$, then Proposition \ref{prop:main} still holds. Thus, we assume the order $5k\gamma^2$ for simplicity in the following theorems.
\end{remark}

For any image $X\in\mathbb{C}^{N\times N}$, its derivatives $X_x$ and $X_y$ belong to $\mathbb{C}^{(N-1)\times N}$ and $\mathbb{C}^{N\times (N-1)}$, respectively. Thus, it is convenient to consider the matrices $\Pi_0$ and $\Pi^0$ obtained from a matrix $\Pi$ by concatenating a row of zeros to the bottom and top of $\Pi$, respectively. More concretely, for a matrix $\Pi\in\mathbb{C}^{(N-1)\times N}$, we denote by $\Pi^0\in\mathbb{C}^{N\times N}$ the augmented matrix with entries
\begin{equation*}
(\Pi^0)_{j,k}=\begin{cases}
0,&j=1,\\
\Pi_{j-1,k},&2\leq j\leq N.
\end{cases}
\end{equation*}
Similarly, we denote by $\Pi_0\in\mathbb{C}^{N\times N}$ the matrix constructed from adding a row of zeros to the bottom of $\Pi$. For a linear operator $\mathcal{A}:\mathbb{C}^{(N-1)\times N}\rightarrow \mathbb{C}^m$ with $[\mathcal{A}(X)]_j=\langle A_j,X\rangle$, we denote by $\mathcal{A}^0:\mathbb{C}^{N\times N}\rightarrow\mathbb{C}^m$ the linear operator with $[\mathcal{A}^0(X)]_j=\langle A_j^0,X\rangle$. We denote by $\mathcal{A}_0:\mathbb{C}^{N\times N}\rightarrow\mathbb{C}^m$ similarly. It was shown in \cite{needell2013stable} that the entire image and its gradients could be related as follows.

\begin{lemma}[\cite{needell2013stable}]\label{lemma:zero}
Given $X\in\mathbb{C}^{N\times N}$ and $\Pi\in\mathbb{C}^{(N-1)\times N}$,
\begin{equation*}
\langle\Pi, X_x \rangle=\langle\Pi^0, X \rangle-\langle\Pi_0, X \rangle\quad\text{and}\quad
\langle\Pi, X_y^{\rm{T}} \rangle=\langle\Pi^0, X^{\rm{T}} \rangle-\langle\Pi_0, X^{\rm{T}} \rangle,
\end{equation*}
where $X^{\rm{T}}$ denotes the (non-conjugate) transpose of the matrix $X$.
\end{lemma}

\subsection{Reconstruction from non-adaptive linear RIP measurements}\label{sec:nonadaptive}
We are prepared to state our first result on stable image reconstruction from non-adaptive linear RIP measurements.

\begin{theorem}\label{thm:1}
Let $N=2^n$ be a power of two, where $n\in\mathbb{N}^+$. Let $\mathcal{A}:\mathbb{C}^{(N-1)\times N}\rightarrow \mathbb{C}^{m_1}$ and $\mathcal{A}':\mathbb{C}^{(N-1)\times N}\rightarrow \mathbb{C}^{m_1}$ be such that the concatenated operator $[\mathcal{A},\mathcal{A}']$ has the RIP of order $5s$ and level $\delta<0.6$. Let ${\mathcal{H}}:\mathbb{C}^{N\times N}\rightarrow\mathbb{C}^{N\times N}$ be the orthonormal bivariate Haar wavelet transform, and ${\mathcal{B}}:\mathbb{C}^{N\times N}\rightarrow \mathbb{C}^{m_2}$ be such that the composite operator ${\mathcal{B}}{\mathcal{H}}^*:\mathbb{C}^{N\times N}\rightarrow\mathbb{C}^{m_2}$ has the RIP of order $2s+1$ and level $\delta<1$. Let $m=4m_1+m_2$, and consider the linear operator ${\mathcal{M}}:\mathbb{C}^{N\times N}\rightarrow\mathbb{C}^m$ with components
\begin{equation}\label{equ:M}
{\mathcal{M}}(X)=\left(\mathcal{A}^0(X),\mathcal{A}_0(X),\mathcal{A}'^0(X^{\rm{T}}),\mathcal{A}'_0(X^{\rm{T}}),{\mathcal{B}}(X)\right).
\end{equation}
Let ${\bar{X}}\in\mathbb{C}^{N\times N}$ be an image and $X^{\rm{opt}}$ the solution to the enhanced TV model \eqref{equ:model} with ${\mathcal{M}}$ defined as \eqref{equ:M}. If $\alpha$ satisfies
\begin{equation}\label{equ:thm1alpha}
\alpha\leq \frac{\sqrt{s}}{2K_2\|\nabla X^{\rm{opt}}\|_2},
\end{equation}
then we have the stable gradient reconstruction results
\begin{equation}\label{equ:gradient1}
\|\nabla \bar{X}-\nabla X^{\rm{opt}}\|_2 \lesssim \sqrt{\frac{\sqrt{s}}{\alpha}\tau+\frac{1}{\alpha}\|\nabla \bar{X}-(\nabla \bar{X})_s\|_1}
\end{equation}
and
\begin{equation}\label{equ:gradient2}
\|\nabla \bar{X}-\nabla X^{\rm{opt}}\|_1 \lesssim \sqrt{s}\sqrt{\frac{\sqrt{s}}{\alpha}\tau+\frac{1}{\alpha}\|\nabla \bar{X}-(\nabla \bar{X})_s\|_1}+\|\nabla \bar{X}-(\nabla \bar{X})_s\|_1,
\end{equation}
and the stable image reconstruction result
\begin{equation}\label{equ:bound1}
\|\bar{X}-X^{\rm{opt}}\|_2 \lesssim \log\left(\frac{N^2}{s}\right)\sqrt{\frac{\sqrt{s}}{\alpha}\tau+\frac{1}{\alpha}\|\nabla \bar{X}-(\nabla \bar{X})_s\|_1}+\log\left(\frac{N^2}{s}\right)\frac{\|\nabla \bar{X}-(\nabla \bar{X})_s\|_1}{\sqrt{s}}+\tau.
\end{equation}
\end{theorem}
\begin{corollary}\label{cor:remove}
Enlightened by Corollary \ref{rem:main}, if
\begin{equation*}
\|\nabla \bar{X}-\nabla X^{\rm{opt}}\|_2 \geq \sqrt{\frac{2}{\alpha}\|\nabla \bar{X}-(\nabla \bar{X})_s\|_1},
\end{equation*}
which is compatible with \eqref{equ:gradient1}, then the linear term $\|\nabla \bar{X}-(\nabla \bar{X})_s\|_1$ in \eqref{equ:gradient2} and hence the term $\log(\frac{N^2}{s})\frac{\|\nabla \bar{X}-(\nabla \bar{X})_s\|_1}{\sqrt{s}}$ in \eqref{equ:bound1} can be removed. This corollary will be proved after Theorem \ref{thm:1}.
\end{corollary}

\begin{remark}\label{rem:technical}
The proof of Theorem \ref{thm:1} is inspired by the proof in \cite{needell2013stable} for the TV model \eqref{equ:TVmodel}, in which it was conjectured that the 4$m_1$ measurements derived from $\mathcal{A}$ in the construction \eqref{equ:M} of ${\mathcal{M}}$ are artifacts of the proof. The components $\mathcal{A}^0(X)$, $\mathcal{A}_0(X)$, $\mathcal{A}'^0(X^{\rm{T}})$, and $\mathcal{A}'_0(X^{\rm{T}})$ are only used for deriving the stable gradient reconstruction bounds \eqref{equ:gradient1} and \eqref{equ:gradient2}. On the other hand, component ${\mathcal{B}}(X)$ only helps us derive the bound \eqref{equ:bound1} from \eqref{equ:gradient1} and \eqref{equ:gradient2}.
\end{remark}

If more measurements are allowed, then the bound \eqref{equ:bound1} can be further improved, the requirement \eqref{equ:thm1alpha} on $\alpha$ can be relaxed, and the artificial components in ${\mathcal{M}}$ can be removed.
\begin{theorem}\label{thm:2}
Let $N=2^n$ be a power of two, where $n\in\mathbb{N}^+$. Let ${\mathcal{H}}:\mathbb{C}^{N\times N}\rightarrow\mathbb{C}^{N\times N}$ be the orthonormal bivariate Haar wavelet transform, and ${\mathcal{M}}:\mathbb{C}^{N\times N}\rightarrow \mathbb{C}^{m}$ be such that the composite operator ${\mathcal{M}}{\mathcal{H}}^*:\mathbb{C}^{N\times N}\rightarrow\mathbb{C}^{m}$ has the RIP of order $Cs\log^3(N)$ and level $\delta<0.6$.
Let ${\bar{X}}\in\mathbb{C}^{N\times N}$ be a mean-zero image or an image containing some zero-valued pixels, and $X^{\rm{opt}}$ be the solution to the enhanced TV model \eqref{equ:model}. If $\alpha$ satisfies
\begin{equation}\label{equ:alpha2}
\alpha \leq \frac{\sqrt{48s\log(N)}}{K_2\|\nabla X^{\rm{opt}}\|_2},
\end{equation}
then we have
\begin{equation}\label{equ:bound2}
\|\bar{X}-X^{\rm{opt}}\|_2\lesssim
\sqrt{\frac{\sqrt{s}}{\alpha}\tau+\frac{1}{\alpha}\|\nabla \bar{X}-(\nabla \bar{X})_{s}\|_1}.
\end{equation}
\end{theorem}

\begin{remark}\label{rem:RIP}
The RIP requirements in both theorems above indicate that the linear measurements should be generated from standard RIP matrix ensembles, which are incoherent with the Haar wavelet system. Many classes of random matrices can be used to generate RIP matrix ensembles. For example, a matrix in $\mathbb{R}^{m\times N^2}$  with i.i.d. normalized Gaussian random entries has a small RIP constant $\delta_s<c$ with high probabilities if $m\gtrsim c^{-2} s\log(N^2/s)$, as shown in \cite{candes2005decoding}. Similar results were extended to sub-Gaussian matrices in \cite{mendelson2007reconstruction}. If $m\gtrsim s\log^4(N)$, then it was proved in \cite{candes2006near,rudelson2008sparse} that the RIP holds with overwhelming probabilities for a partial Fourier matrix ${\mathcal{F}}_{\Omega}\in\mathbb{R}^{m\times N^2}$. The RIP also holds for randomly generated circulant matrices (see \cite{rauhut2012restricted}) and randomly subsampled bounded orthonormal systems (see \cite{rauhut2012sparse}). Most of these mentioned measurements are incoherent with the Haar wavelet system, but the partial Fourier matrix with uniformly subsampled rows is an exception. Thus, some specific sampling strategies for Fourier measurements should be considered. For example, it was asserted in \cite{krahmer2011new} that ${\mathcal{F}}_{\Omega}\in\mathbb{R}^{m\times N^2}$ with $m\gtrsim s\log^4(N)$ and \emph{randomized column signs} has the RIP; it was also shown in \cite{krahmer2013stable} that ${\mathcal{F}}_{\Omega}$ with rows subsampled according to some power-law densities is incoherent with the Haar wavelet system after preconditioning.
\end{remark}

\subsection{Reconstruction from variable-density Fourier measurements}\label{sec:fourier}
As shown in \cite{krahmer2013stable}, if the measurements are sampled according to appropriate power-law densities, then they are incoherent with the Haar wavelet system. We consider a particular variable-density sampling strategy proposed in \cite{krahmer2013stable} and derive a partial stable image reconstruction theorem tailored for Fourier measurements. Following the idea of \cite{krahmer2013stable}, our guarantees are based on a \emph{weighted $\ell_2$-norm} in measuring noise such that high-frequency measurements have a higher sensitivity to noise; that is, the $\ell_2$-norm in the constraint $\|{\mathcal{M}} X-y\|_2\leq \tau$ of the enhanced TV model \eqref{equ:model} is replaced by a weighted $\ell_2$-norm model. For the particular scenario with Fourier measurements, the general linear operator ${\mathcal{M}}$ is specified as ${\mathcal{F}}_{\Omega}$, which is the restriction of the Fourier transform matrix to a set $\Omega$ of frequencies as defined in Section \ref{sec:fouriersystem}.

\begin{theorem}\label{thm:3}
Let $N = 2^n$ be a power of 2, where $n\in\mathbb{N}^+$. Let $m$ and $s$ satisfy $s\gtrsim \log(N)$ and
\begin{equation}\label{equ:m}
m\gtrsim s\log^3(s)\log^5(N).
\end{equation}
Select $m$ frequencies $\{(\omega^j_1,\omega^j_2)\}_{j=1}^m\subset \{-N/1+2,\ldots,N/2\}^2$ i.i.d. according to
\begin{equation}\label{equ:sampling}
\mathbb{P}[(\omega^j_1,\omega^j_2)=(k_1,k_2)]=C_N\min\left(C,\frac{1}{k_1^2+k_2^2}\right)=:\eta(k_1,k_2),\quad -N/2+1\leq k_1,k_2\leq N/2,
\end{equation}
where $C$ is an absolute constant and $C_N$ is chosen such that $\eta$ is a probability distribution. Consider the weight vector $\rho=(\rho_j)_{j=1}^m$ with $\rho_j=[1/\eta(\omega^j_1,\omega^j_2)]^{1/2}$. Then we have the following assertion for all mean-zero or zero-valued pixel-containing images $\bar{X}\in\mathbb{C}^{N\times N}$ with probability exceeding $1-N^{-C\log^3(s)}$: Given noisy partial Fourier measurements $b={\mathcal{F}}_{\Omega}\bar{X}+e$, if
\begin{equation}\label{equ:alpha3}
\alpha \leq \frac{\sqrt{48s\log(N)}}{K_2\|\nabla X^{\rm{opt}}\|_2},
\end{equation}
then the solution $X^{\rm{opt}}$ to the model
\begin{equation}\label{equ:fouriermodel}
\min_{X\in\mathbb{C}^{N\times N}}~\|\nabla X\|_1-\frac{\alpha}{2}\|\nabla X\|_2^2\quad{\rm{s.t.}}\quad \|\rho\circ({\mathcal{F}}_{\Omega} X-b)\|_2\leq\tau\sqrt{m}
\end{equation}
satisfies
\begin{equation}\label{equ:bound3}
\|\bar{X}-X^{\rm{opt}}\|_2\lesssim
\sqrt{\frac{\sqrt{s}}{\alpha}\tau+\frac{1}{\alpha}\|\nabla \bar{X}-(\nabla \bar{X})_{s}\|_1}.
\end{equation}
\end{theorem}

\subsection{Further discussion}\label{sec:discussion}

We supplement more details about the theoretical results presented in Sections \ref{sec:nonadaptive} and \ref{sec:fourier}.

\noindent \textbf{The \emph{a posterior} verification on $\alpha$.} Three conditions \eqref{equ:thm1alpha}, \eqref{equ:alpha2}, and \eqref{equ:alpha3} on $\alpha$ are required in Theorems \ref{thm:1}, \ref{thm:2}, and \ref{thm:3}, respectively. Determining the value of $\alpha$ is possible only if we have \emph{a priori} estimation on $\|X^{\rm{opt}}\|_2$. Thus, these conditions can be interpreted as \emph{a posterior} verification because they can be verified once $X^{\rm{opt}}$ is obtained by solving the model \eqref{equ:model}. In practice, we solve the model \eqref{equ:model} numerically and thus obtain an approximate solution, denoted by $X^*$, subject to a preset accuracy $\epsilon>0$. That is, $\|X^{\rm{opt}}-X^*\|_2\leq\epsilon$. Then, if
\begin{equation*}
\alpha\leq \frac{\sqrt{s}}{2K_2(\|\nabla X^{*}\|_2+\epsilon)},
\end{equation*}
then \eqref{equ:thm1alpha} is guaranteed; if
\begin{equation*}
\alpha\leq  \frac{\sqrt{48s\log(N)}}{K_2(\|\nabla X^{*}\|_2+\epsilon)},
\end{equation*}
then \eqref{equ:alpha2} and \eqref{equ:alpha3} are satisfied.

\noindent \textbf{The RIP level $\delta<0.6$ in Theorems \ref{thm:1} and \ref{thm:2}.} The bound 0.6 is sharp, as we need to ensure $\sqrt{1-\delta}-\sqrt{1+\delta}/2>0$ (cf. proof in Section \ref{sec:proposition}). For the reconstruction guarantees  derived in \cite{needell2013stable} for the TV model \eqref{equ:TVmodel}, the level is assumed to satisfy $\delta<1/3$, and it is not sharp as remarked in \cite{needell2013stable}. Though $\delta<1/3$ can be improved, the reconstruction error bound in \cite{needell2013stable} for the TV model \eqref{equ:TVmodel} tends to be infinity if $\delta\rightarrow 0.6$. In light of Remark \ref{rem:reasonable}, the bounds \eqref{equ:bound1} and \eqref{equ:bound2} are still valid in this case, and the upper bound required for $\alpha$ tends to $0$ correspondingly with consideration of the behavior of $K_2$. That is, Theorems \ref{thm:1} and \ref{thm:2} can guarantee the stability of the TV model \eqref{equ:TVmodel} when $\delta\rightarrow0.6$, resulting in reconstruction error bounds in forms of \eqref{equ:bound1} and \eqref{equ:bound2}.

\noindent \textbf{The required amount $m$ of Fourier measurements in Theorem \ref{thm:3}.} The RIP level $\delta$ does not appear explicitly in Theorem \ref{thm:3}, while we shall assume $m\gtrsim s\delta^{-2}\log^3(s)\log^5(N)$ and the constant $\delta$ is eliminated in such an inequality with $\gtrsim$; see our proof in Section \ref{sec:proof3}. The least required amount $m$ for the TV model \eqref{equ:TVmodel} shall also satisfy this relation with $s$, $N$, and $\delta$, as proved in \cite{krahmer2013stable}. Since the upper bound on the RIP level $\delta$ is enlarged from $1/3$ for the TV model \eqref{equ:TVmodel} (see \cite{krahmer2013stable}) to 0.6 for the enhanced TV model \eqref{equ:model}, the least amount of Fourier measurements required for the enhanced TV model \eqref{equ:model} should be $\left(0.6/(1/3)\right)^{-2}\approx 30.86\%$ of the least amount of Fourier measurements required in \cite{krahmer2013stable} for the TV model \eqref{equ:TVmodel}.

\noindent \textbf{Inconsistency when $\alpha\rightarrow0$.} The enhanced TV regularization \eqref{equ:penalty} tends to be the anisotropic TV term as $\alpha\rightarrow0$. At the same time, the reconstruction error bounds \eqref{equ:bound1}, \eqref{equ:bound2}, and \eqref{equ:bound3} do not reduce to the corresponding bounds \eqref{equ:TVrecovery} and \eqref{equ:TVrecoveryimproved} for the TV model \eqref{equ:TVmodel}. Note that the bounds \eqref{equ:bound2} and \eqref{equ:bound3} are of the same form. To explain this inconsistency, note that Proposition \ref{prop:main} is a pillar of the proofs of Theorems \ref{thm:1}, \ref{thm:2}, and \ref{thm:3}. In contrast, the proof for the TV model \eqref{equ:TVmodel} in \cite{needell2013stable} relies on the following fact: If $D$ satisfies the tube constraint \eqref{equ:tube} and the cone constraint $\|D_{{S}^c}\|_1\leq \gamma\|D_{S}\|_1+\sigma$,
then it was shown in \cite{needell2013stable} that
\begin{equation}\label{equ:D1TV}
\|D\|_2\lesssim \frac{\sigma}{\gamma\sqrt{k}}+\varepsilon\quad\text{and}\quad
\|D\|_1\lesssim \sigma+{\gamma\sqrt{k}}\varepsilon.
\end{equation}
Indeed, the left-hand side of the estimation \eqref{equ:inequ} in the proof of Proposition \ref{prop:main} contains a quadratic term $\|D\|_2^2$ and a linear term $\|D\|_2$, and only the linear term remains if $\beta_1,\beta_2\rightarrow0$, which then leads to the same result as \eqref{equ:D1TV}. However, in the proof of Proposition \ref{prop:main}, we remove this linear term and keep the quadratic term, and hence the obtained result cannot be reduced to the result \eqref{equ:D1TV} as $\beta_1,\beta_2\rightarrow0$. Such an inconsistent situation is also encountered by the springback model in \cite{an2021springback}.

\noindent \textbf{Comparison between \eqref{equ:TVrecoveryimproved} and \eqref{equ:bound2}.} We are interested in whether or not the bound \eqref{equ:bound2} (as well as the bound \eqref{equ:bound3}, which shares the same form as \eqref{equ:bound2}) can be tighter than \eqref{equ:TVrecoveryimproved} in the sense of
\begin{equation}\label{equ:comp}
\sqrt{\frac{\sqrt{s}}{\alpha}\tau+\frac{1}{\alpha}\|\nabla \bar{X}-(\nabla \bar{X})_{s}\|_1}\lesssim \frac{\|\nabla {\bar{X}} -(\nabla {\bar{X}})_s\|_1}{\sqrt{s}}+\tau,
\end{equation}
with a given $\alpha>0$. If the image ${\bar{X}}$ is known to have an $s$-sparse gradient, then the comparison \eqref{equ:comp} is reduced to $\sqrt{s}\lesssim \alpha\tau$. As $s$ is fixed in this scenario, we can claim that the estimation \eqref{equ:bound2} is tighter than the estimation \eqref{equ:TVrecoveryimproved} in the sense of \eqref{equ:comp} if $\tau\gtrsim \sqrt{s}/\alpha$, i.e., the level of noise $\tau$ is \emph{relatively large}. If the sparsity of $\nabla {\bar{X}}$ is not assumed, but the linear measurements are noise-free, i.e., $\tau=0$, then the comparison \eqref{equ:comp} is reduced to
\begin{equation}\label{equ:nosparsebutnoisefree}
s/\|\nabla \bar{X}-(\nabla \bar{X})_{s}\|_1\lesssim \alpha,
\end{equation}
in which the left-hand side of \eqref{equ:nosparsebutnoisefree} is an increasing function of $s$. In order to discern the scenario where \eqref{equ:nosparsebutnoisefree} holds, a key fact from Remark \ref{rem:RIP} should be noticed: for RIP measurements mentioned there, a small number $m$ of measurements admits an RIP with a small $s$. The bound $\mathcal{O}(s\log(N^2/s))$ for Gaussian measurements appears not to be monotonic with respect to $s$. On the other hand, with the implicit constant factors derived in \cite{rudelson2008sparse}, this bound is indeed monotonically increasing with respect to $s$. Thus, if \emph{the number of measurements $m$ is limited}, which only renders an RIP with a small $s$, then \eqref{equ:nosparsebutnoisefree} holds. This situation coincides with the intuition that, as the term $\|\nabla \bar{X}-(\nabla \bar{X})_{s}\|_1\gg1$ for many digital images, especially when the number of measurements is limited (so that $s$ is small), taking a square root shall lead to a smaller bound than that without doing so. 

Together with both scenarios, we can claim that if the level of noise $\tau$ is \emph{relatively large} and \emph{the number of measurements $m$ is limited}, then the enhanced TV model \eqref{equ:model} performs better than the TV model \eqref{equ:TVmodel} in the sense of \eqref{equ:comp}, because \eqref{equ:comp} is guaranteed to hold when
\begin{equation*}
\sqrt{\frac{\sqrt{s}}{\alpha}\tau}+\sqrt{\frac{1}{\alpha}\|\nabla \bar{X}-(\nabla \bar{X})_{s}\|_1}\lesssim \frac{\|\nabla {\bar{X}} -(\nabla {\bar{X}})_s\|_1}{\sqrt{s}}+\tau,
\end{equation*}
and we can study $\sqrt{\frac{\sqrt{s}}{\alpha}\tau}\lesssim \tau$ and $\sqrt{\frac{1}{\alpha}\|\nabla \bar{X}-(\nabla \bar{X})_{s}\|_1}\lesssim\frac{\|\nabla {\bar{X}} -(\nabla {\bar{X}})_s\|_1}{\sqrt{s}}$ separately.

This comparison can be analogously extended to other cases for which the corresponding reconstruction error bounds are also linear with respect to terms $\|\nabla {\bar{X}} -(\nabla {\bar{X}})_s\|_1/\sqrt{s}$ and $\tau$. Such examples include the model in \cite{lou2015weighted}, which has the regularization term $\|X\|_{\text{TV}_a}-\|X\|_{\text{TV}_i}$. For the model in \cite{lou2015weighted}, it seems that reconstruction guarantees leading to an error bound without the log factor $\log(N^2/s)$ are still missing. Note that this log factor also occurs in the bound \eqref{equ:TVrecovery} for the TV model \eqref{equ:TVmodel} and the bound \eqref{equ:bound1} for the enhanced TV model \eqref{equ:model}, but it is removed if the required RIP order increases from $\mathcal{O}(s)$ to $\mathcal{O}(s\log^3(N))$, and then both bounds can be improved to \eqref{equ:TVrecoveryimproved} and \eqref{equ:bound2}, respectively. Reconstruction guarantees for the model in \cite{lou2015weighted} have been investigated in \cite{li2020ell1}. However, the derived error bound (see Theorem 3.8 in \cite{li2020ell1}) still fails to remove the log factor $\log(N^2/s)$, despite that the subsampled measurements are required to have the RIP of order $\mathcal{O}(s^2\log(N))$ with a more complicated level $\delta$ which depends on $N$, $s$, and the constant $\tilde{C}$ in Lemma \ref{lem:1}.

\section{Proofs}\label{sec:guarantee}

In this section, we present the complete proofs for the theoretical results in Section \ref{sec:theory}.

\subsection{Proofs of Proposition \ref{prop:main} and Corollary \ref{rem:main}}\label{sec:proposition}
\noindent \emph{Proof of Proposition \ref{prop:main}.} We arrange the indices in $S^c$ in order of decreasing magnitudes (in absolute value) of $D_{S^c}$ and divide $S^c$ into subsets of size $4k\gamma^2$, i.e., $S^c=S_1\bigcup S_2\bigcup\cdots\bigcup S_{r}$, where $r=\left\lfloor\frac{N^2-|S|}{4k\gamma^2}\right\rfloor$. In other words, $D_{S^c}=D_{S_1}+D_{S_2}+\cdots+D_{S_r}$, where $D_{S_1}$ consists of the $4k\gamma^2$ largest-magnitude components of $D$ over $S^c$, $D_{S_2}$ consists of the next $4k\gamma^2$ largest-magnitude components of $D$ over $S^c\backslash S_1$, and so forth. As the magnitude of each component of $D_{S_{j}}$ is less than the average magnitude $\|D_{S_{j-1}}\|_1/(4k\gamma^2)$ of components of $D_{S_{j-1}}$, we have
\begin{equation*}
\|D_{S_j}\|_2^2\leq 4k\gamma^2 \left(\frac{\|D_{S_{j-1}}\|_1}{4k\gamma^2}\right)^2=\frac{\|D_{S_{j-1}}\|_1^2}{4k\gamma^2},\quad j = 2,3,\ldots,r.
\end{equation*}
Thus, combining $\|D_{S_j}\|_2\leq \frac{\|D_{S_{j-1}}\|_1}{2\gamma\sqrt{k}}$ with the cone constraint \eqref{equ:cone}, we have
\begin{equation*}
\sum_{j=2}^r\|D_{S_j}\|_2\leq \frac{1}{2\gamma\sqrt{k}}\|D_{S^c}\|_1\leq \frac{\|D_S\|_1}{2\sqrt{k}}-\frac{\beta_1}{4\gamma\sqrt{k}}\|D\|_2^2+\frac{\sigma}{2\gamma\sqrt{k}}+\frac{\beta_2}{2\gamma\sqrt{k}}\left<E_1,E_2\right>.
\end{equation*}
The assumption $|S|\leq k$ leads to $\|D_S\|_1\leq \sqrt{|S|}\|D_S\|_2\leq\sqrt{k}\|D_S\|_2\leq \sqrt{k}\|D_S+D_{S_1}\|_2$,
hence we have
\begin{equation}\label{equ:scbound}
\sum_{j=2}^r\|D_{S_j}\|_2\leq  \frac{\|D_S+D_{S_1}\|_2}{2}-\frac{\beta_1}{4\gamma\sqrt{k}}\|D\|_2^2+\frac{\sigma}{2\gamma\sqrt{k}}+\frac{\beta_2}{2\gamma\sqrt{k}}\left<E_1,E_2\right>.
\end{equation}
Together with this bound \eqref{equ:scbound}, the tube constraint \eqref{equ:tube}, and the RIP of $\mathcal{A}$, we have
\begin{equation*}\begin{split}
\varepsilon &\geq \|\mathcal{A} D\|_2 \geq \|\mathcal{A} (D_S+D_{S_1})\|_2-\sum_{j=2}^r\|\mathcal{A} D_{S_j}\|_2  \geq \sqrt{1-\delta}\|D_S+D_{S_1}\|_2-\sqrt{1+\delta} \sum_{j=2}^r\| D_{S_j}\|_2\\
& \geq \sqrt{1-\delta}\|D_S+D_{S_1}\|_2-\sqrt{1+\delta} \left(\frac{\|D_S+D_{S_1}\|_2}{2}-\frac{\beta_1}{4\gamma\sqrt{k}}\|D\|_2^2+\frac{\sigma}{2\gamma\sqrt{k}}+\frac{\beta_2}{2\gamma\sqrt{k}}\left<E_1,E_2\right>\right)\\
& =\left(\sqrt{1-\delta}-\frac{\sqrt{1+\delta}}{2}\right)\|D_S+D_{S_1}\|_2
+\frac{\beta_1\sqrt{1+\delta}}{4\gamma\sqrt{k}}\|D\|_2^2-\frac{\sqrt{1+\delta}}{2\gamma\sqrt{k}}\sigma-\frac{\beta_2\sqrt{1+\delta}}{2\gamma\sqrt{k}}\left<E_1,E_2\right>.
\end{split}\end{equation*}
The assumption $\delta<0.6$ ensures $\sqrt{1-\delta}-\sqrt{1+\delta}/{2}>0$. Hence, we have
\begin{equation*}
\|D_S+D_{S_1}\|_2\leq\frac{2}{2\sqrt{1-\delta}-\sqrt{1+\delta}}\left(\varepsilon -\frac{\beta_1\sqrt{1+\delta}}{4\gamma\sqrt{k}}\|D\|_2^2+\frac{\sqrt{1+\delta}}{2\gamma\sqrt{k}}\sigma+\frac{\beta_2\sqrt{1+\delta}}{2\gamma\sqrt{k}}\left<E_1,E_2\right>     \right).
\end{equation*}
As $\|D\|_2$ is bounded by the sum of $\|D_S+D_{S_1}\|_2$ and $\sum_{j=2}^r\|D_{S_j}\|_2$, it satisfies
\begin{equation*}\begin{split}
\|D\|_2
\leq& \frac{3}{2}\|D_S+D_{S_1}\|_2 -\frac{\beta_1}{4\gamma\sqrt{k}}\|D\|_2^2+\frac{\sigma}{2\gamma\sqrt{k}}+\frac{\beta_2}{2\gamma\sqrt{k}}\left<E_1,E_2\right>\\
\leq&\frac{3}{2\sqrt{1-\delta}-\sqrt{1+\delta}}\varepsilon +\left(\frac{3}{2\sqrt{1-\delta}-\sqrt{1+\delta}}+\frac{1}{\sqrt{1+\delta}}\right)\cdot\\
&\left(-\frac{\beta_1\sqrt{1+\delta}}{4\gamma\sqrt{k}}\|D\|_2^2+\frac{\sqrt{1+\delta}}{2\gamma\sqrt{k}}\sigma+\frac{\beta_2\sqrt{1+\delta}}{2\gamma\sqrt{k}}\left<E_1,E_2\right> \right)\\
:=& K_1\varepsilon-\frac{\beta_1 K_2}{\gamma\sqrt{k}}\|D\|_2^2+\frac{2 K_2}{\gamma\sqrt{k}}\sigma+\frac{2\beta_2 K_2}{\gamma\sqrt{k}}\left<E_1,E_2\right>.
\end{split}\end{equation*}
Thus, we have the quadratic inequality
\begin{equation}\label{equ:inequ}
\frac{\beta_1 K_2}{\gamma\sqrt{k}}\|D\|_2^2+\|D\|_2-\frac{2\beta_2 K_2}{\gamma\sqrt{k}}\left<E_1,E_2\right>-K_1\varepsilon-\frac{2 K_2}{\gamma\sqrt{k}}\sigma\leq 0.
\end{equation}
The requirement \eqref{equ:alpha} on $\beta_2$ ensures that
$$\|D\|_2-\frac{2\beta_2 K_2}{\gamma\sqrt{k}}\left<E_1,E_2\right>\geq \|D\|_2-\left<\frac{E_1}{\|E_1\|_2},E_2\right>\geq 0,$$
where the last inequality is due to Cauchy--Schwarz inequality and $\|E_2\|_2=\|D\|_2$. Then, we have
$$\frac{\beta_1 K_2}{\gamma\sqrt{k}}\|D\|_2^2-K_1\varepsilon-\frac{2 K_2}{\gamma\sqrt{k}}\sigma\leq 0,$$
which yields the estimation \eqref{equ:D2}. Finally, we derive \eqref{equ:D1D2}. As $|S|\leq k$, we have $\|D_S\|_1\leq\sqrt{k}\|D_S\|_2$. Then, together with the requirement \eqref{equ:alpha} on $\beta_2$ and the cone constraint \eqref{equ:cone}, we have
\begin{equation}\label{equ:DDD}\begin{split}
\|D\|_1\leq&(\gamma+1)\|D_S\|_1-\frac{\beta_1}{2}\|D\|_2^2+\sigma+\beta_2\left<E_1,E_2\right>\leq (\gamma+1)\|D_S\|_1+\sigma+\frac{\gamma\sqrt{k}}{2K_2}\|D\|_2\\
\leq &(\gamma+1)\sqrt{k}\|D_S\|_2+\sigma+\frac{\gamma\sqrt{k}}{2K_2}\|D\|_2\leq (\gamma+1)\sqrt{k}\|D\|_2+\sigma+\frac{\gamma\sqrt{k}}{2K_2}\|D\|_2\\
=&\frac{(2K_2+1)\gamma\sqrt{k}+2K_2\sqrt{k}}{2K_2}\|D\|_2+\sigma,
\end{split}\end{equation}
which completes the proof of Proposition \ref{prop:main}. \hfill $\square$

\noindent \emph{Proof of Corollary \ref{rem:main}.} In the second inequality of \eqref{equ:DDD}, we use the fact $-\frac{\beta_1}{2}\|D\|_2^2\leq0$. If $\|D\|_2$ satisfies $\|D\|_2\geq\sqrt{2\sigma/\beta_1}$, then $-\frac{\beta_1}{2}\|D\|_2^2+\sigma\leq 0$ and it follows from \eqref{equ:DDD} that
\begin{equation*}\begin{split}
\|D\|_1\leq&(\gamma+1)\|D_S\|_1-\frac{\beta_1}{2}\|D\|_2^2+\sigma+\beta_2\left<E_1,E_2\right>\\
\leq& (\gamma+1)\|D_S\|_1+\frac{\gamma\sqrt{k}}{2K_2}\|D\|_2\leq\frac{(2K_2+1)\gamma\sqrt{k}+2K_2\sqrt{k}}{2K_2}\|D\|_2,
\end{split}\end{equation*}
which completes the proof of Corollary \ref{rem:main}.  \hfill $\square$

\subsection{Proof of Theorem \ref{thm:1} and Corollary \ref{cor:remove}}\label{proof:2}
We first prove the stable gradient reconstruction results \eqref{equ:gradient1} and \eqref{equ:gradient2}, and then obtain the stable image reconstruction result \eqref{equ:bound1} with the aid of a strong Sobolev inequality. The following Sobolev inequality was derived in \cite{needell2013stable} for images with multivariate generalization given in \cite{needell2013near}.
\begin{lemma}[Strong Sobolev inequality]\label{lem:sobolev}
Let ${\mathcal{B}}:\mathbb{C}^{N\times N}\rightarrow\mathbb{C}^m$ be a linear map such that ${\mathcal{B}}{\mathcal{H}}^*:\mathbb{C}^{N\times N}\rightarrow\mathbb{C}^m$ has the RIP of order $2s+1$ and level $\delta<1$, where ${\mathcal{H}}:\mathbb{C}^{N\times N}\rightarrow\mathbb{C}^{N\times N}$ is the bivariate Haar transform. Suppose that $D\in\mathbb{C}^{N\times N}$ satisfies the tube constraint $\|{\mathcal{B}} D\|_2\leq\varepsilon$. Then
$$\|D\|_2\leq C_2 \left[ \left(\frac{\|\nabla D\|_{1}}{\sqrt{s}}\right)\log\left(\frac{N^2}{s}\right)+\varepsilon\right].$$
\end{lemma}

\noindent \emph{Proof of Theorem \ref{thm:1}.} The proof is divided into the stable gradient and image reconstructions, respectively.

\noindent\textbf{Stable gradient reconstruction.} We plan to apply Proposition \ref{prop:main} to the term $\nabla(X^{\text{opt}}-\bar{X})$. Let $V=X^{\text{opt}}-\bar{X}$ and $L=(V_x,V_y^{\text{T}})$. For convenience, let $P$ denote the mapping of indices which maps the index of a nonzero entry in $\nabla V$ to its corresponding index in $L$. By the definition of $\nabla$, $L$ has the same norm as $\nabla V$, i.e., $\|L\|_2=\|\nabla V\|_2$ and $\|L\|_1=\|\nabla V\|_1$. Thus, it suffices to apply Proposition \ref{prop:main} to $L$. Let $A_1,A_2,\ldots,A_{m_1},A'_1,A'_2,\ldots,A'_{m_1}$ be such that
$[\mathcal{A}(Z)]_j=\left<A_j,Z\right>$ and $[\mathcal{A}'(Z)]_j=\left<A'_j,Z\right>$.
\begin{itemize}
       \item \emph{Cone constraint.} Let $S$ denote the support of the largest $s$ entries of $\nabla \bar{X}$. On one hand, it holds that
$$\|\nabla X^{\rm{opt}}\|_1-\frac{\alpha}{2}\|\nabla X^{\rm{opt}}\|_{2}^2\leq \|\nabla \bar{X}\|_{1}-\frac{\alpha}{2}\|\nabla \bar{X}\|_{2}^2
= \|(\nabla \bar{X})_S\|_1+\|(\nabla \bar{X})_{S^c}\|_1-\frac{\alpha}{2}\|\nabla\bar{X}\|_2^2.$$
On the other hand, we have
\begin{equation*}\begin{split}
&\|\nabla X^{\rm{opt}}\|_{1}-\frac{\alpha}{2}\|\nabla X^{\rm{opt}}\|_{2}^2\\
=&\|(\nabla \bar{X})_S+(\nabla V)_S\|_1+\|(\nabla \bar{X})_{S^c}+(\nabla V)_{S^c}\|_1-\frac{\alpha}{2}\|\nabla\bar{X}+\nabla V\|_2^2\\
\geq & \|(\nabla \bar{X})_S\|_1-\|(\nabla V)_S\|_1+\|(\nabla V)_{S^c}\|_1- \|(\nabla \bar{X})_{S^c}\|_1-\frac{\alpha}{2}\left(\|\nabla\bar{X}\|_2^2+2\left<\nabla\bar{X},\nabla V\right>+\|\nabla V\|_2^2\right).
\end{split}\end{equation*}
Thus, we obtain
\begin{equation*}\begin{split}
\|(\nabla V)_{S^c}\|_1\leq &\|(\nabla V)_S\|_1+2\|(\nabla \bar{X})_{S^c}\|_1+\frac{\alpha}{2}\|\nabla V\|_2^2+\alpha\left<\nabla\bar{X},\nabla V\right>\\
=&\|(\nabla V)_S\|_1+2\|\nabla \bar{X}-(\nabla \bar{X})_{s}\|_1-\frac{\alpha}{2}\|\nabla V\|_2^2+\alpha\left<\nabla X^{\rm{opt}},\nabla V\right>.
\end{split}\end{equation*}
As $L$ contains all the same nonzero entries as $\nabla V$, it satisfies the following cone constraint:
$$\|L_{P(S)^c}\|_1\leq \|L_{P(S)}\|_1+2\|\nabla \bar{X}-(\nabla \bar{X})_{s}\|_1-\frac{\alpha}{2}\|L\|_2^2+\alpha\left<\nabla X^{\rm{opt}},\nabla V\right>.$$
\item \emph{Tube constraint.} We note that $V$ satisfies a tube constraint as
$$\|{\mathcal{M}} V\|_2^2=\|({\mathcal{M}} X^{\rm{opt}}-y)-({\mathcal{M}}\bar{X}-y)\|_2^2\leq2\|{\mathcal{M}} X^{\rm{opt}}-y\|_2^2+2\|{\mathcal{M}}\bar{X}-y\|_2^2\leq 4\tau^2.$$
Then, it follows from Lemma \ref{lemma:zero} that
$$|\left<A_j,V_x\right>|^2=|\left<[A_j]^0,V\right>-\left<[A_j]_0,V\right>|^2\leq 2|\left<[A_j]^0,V\right>|^2+2|\left<[A_j]_0,V\right>|^2$$
and
$$|\left<A'_j,V^{\rm{T}}_y\right>|^2=|\left<[A_j']^0,V^{\rm{T}}\right>-\left<[A'_j]_0,V^{\rm{T}}\right>|^2\leq 2|\left<[A_j']^0,V^{\rm{T}}\right>|^2+2|\left<[A'_j]_0,V^{\rm{T}}\right>|^2.$$
Thus, $L$ also satisfies a tube constraint:
$$\|[\mathcal{A}~\mathcal{A}']L\|_2^2=\sum_{j=1}^m|\left<A_j,V_x\right>|^2+|\left<A'_j,V^{\rm{T}}_y\right>|^2\leq 2\|{\mathcal{M}}(V)\|_2^2\leq 8 \tau^2.$$

By virtue of Proposition \ref{prop:main} with $\gamma=1$, $k=s$, $\beta_1=\beta_2=\alpha$, $\sigma = 2\|\nabla \bar{X}-(\nabla \bar{X})_{s}\|_1$, $\varepsilon=2\sqrt{2}\tau$, $E_1=\nabla X^{\rm{opt}}$ and $E_2=\nabla V$, the requirement \eqref{equ:thm1alpha} of $\alpha$ ensures that
\begin{equation*}\begin{split}
\|\nabla X^{\text{opt}}-\nabla\bar{X}\|_2=\|L\|_2\leq \sqrt{\frac{2\sqrt2\sqrt{s}K_1}{\alpha K_2}\tau+\frac{4}{\alpha}\|\nabla X-(\nabla X)_{s}\|_1}.
\end{split}\end{equation*}
Furthermore, by \eqref{equ:D1D2}, we have $\|\nabla X^{\text{opt}}-\nabla\bar{X}\|_1=\|L\|_1$ and
\begin{equation}\label{equ:forremoving}\begin{split}
\|L\|_1
\leq\frac{(4K_2+1)\sqrt{s}}{2K_2}\sqrt{\frac{2\sqrt2\sqrt{s}K_1}{\alpha K_2}\tau+\frac{4}{\alpha}\|\nabla \bar{X}-(\nabla \bar{X})_{s}\|_1}+2\|\nabla \bar{X}-(\nabla \bar{X})_{s}\|_1,
\end{split}\end{equation}
which completes the proof of the stable gradient reconstruction results \eqref{equ:gradient1} and \eqref{equ:gradient2}.
\end{itemize}

\noindent\textbf{Stable image reconstruction.} We now apply the strong Sobolev inequality given in Lemma \ref{lem:sobolev} to $X^{\rm{opt}}-\bar{X}$. As $\|{\mathcal{B}}(X^{\rm{opt}}-\bar{X})\|_2\leq\|{\mathcal{M}}(X^{\rm{opt}}-\bar{X})\|_2\leq2\tau$, we have
$$\|X^{\rm{opt}}-\bar{X}\|_2\lesssim \log\left(\frac{N^2}{s}\right)\left(\frac{\|\nabla X^{\rm{opt}}-\nabla \bar{X}\|_{1}}{\sqrt{s}}\right)+\tau.$$
Together with the bound \eqref{equ:gradient2}, we have the stable image reconstruction result \eqref{equ:bound1}. \hfill $\square$

\noindent \emph{Proof of Corollary \ref{cor:remove}.} If $\|\nabla \bar{X}-\nabla X^{\rm{opt}}\|_2 \geq \sqrt{\frac{2}{\alpha}\|\nabla \bar{X}-(\nabla \bar{X})_s\|_1}$, then it follows from Corollary \ref{rem:main} that the linear term of $\|\nabla \bar{X}-(\nabla \bar{X})_{s}\|_1$ in the estimation \eqref{equ:forremoving} can be removed. Thus, from \eqref{equ:forremoving} to \eqref{equ:bound1}, the term $\log(\frac{N^2}{s})\frac{\|\nabla \bar{X}-(\nabla \bar{X})_s\|_1}{\sqrt{s}}$ in \eqref{equ:bound1} can be also removed. \hfill $\square$

\subsection{Proof of Theorem \ref{thm:2}}
We apply Proposition \ref{prop:main} to $c={\mathcal{H}} V$ as opposed to $\nabla V$. Some properties of the bivariate Haar wavelet system, characterized as Lemmas \ref{lem:1}, \ref{lem:2}, and \ref{lem:2}, are needed in the proof. Besides, a classical Sobolev inequality weaker than the strong Sobolev inequality in Lemma \ref{lem:sobolev} is needed.
\begin{lemma}[\cite{needell2013stable}]
Let $X\in\mathbb{C}^{N\times N}$ be a mean-zero image or contain some zero-valued pixels. Then
\begin{equation}\label{equ:classicalsobolev}
\|X\|_2\leq\|\nabla X\|_1.
\end{equation}
\end{lemma}

\noindent \emph{Proof of Theorem \ref{thm:2}.} Let $V=X^{\text{opt}}-\bar{X}$, and apply Proposition \ref{prop:main} to $c={\mathcal{H}} V$, where $c_{(1)}:=c_{(1)}(V)$ denotes the Haar coefficient corresponding to the constant wavelet, and $c_{(j)}:=c_{(j)}(V)$ ($j\geq2$) denotes the $(j-1)$-st largest-magnitude Haar coefficient among the remaining. We use this ordering because Lemma \ref{lem:1} applies only to mean-zero images. Let $h_{(j)}$ denote the Haar wavelet associated with $c_{(j)}$. We have assumed that the composite operator ${\mathcal{M}}{\mathcal{H}}^*:\mathbb{C}^{N\times N}\rightarrow\mathbb{C}^{m}$ has the RIP of order $Cs\log^3(N)$ and level $\delta<0.6$, and we now derive the constant $C$.
\begin{itemize}
       \item \emph{Cone constraint on $c={\mathcal{H}} V$.} As shown in Section \ref{proof:2}, we have
\begin{equation}\label{equ:cone3}
\|(\nabla V)_{S^c}\|_1\leq \|(\nabla V)_S\|_1+2\|\nabla \bar{X}-(\nabla \bar{X})_{s}\|_1-\frac{\alpha}{2}\|\nabla V\|_2^2+\alpha\left<\nabla X^{\rm{opt}},\nabla V\right>.
\end{equation}
Recall that $S$ is the index set of $s$ largest-magnitude entries of $\nabla V$. It follows from Lemma \ref{lem:2} that the set $\Omega$ of wavelets which are non-constant over $S$ has the cardinality at most $6s\log(N)$, i.e., $|\Omega|\leq6s\log(N)$. Decompose $V$ as
$$V=\sum_j c_{(j)}h_{(j)}=\sum_{j\in\Omega}c_{(j)}h_{(j)}+\sum_{j\in\Omega^c}c_{(j)}h_{(j)}=: V_{\Omega}+V_{\Omega^c}.$$
Because of the linearity of $\nabla$, we have $\nabla V = \nabla V_{\Omega}+\nabla V_{\Omega^c}$. By the construction of $\Omega$, we have $(\nabla V_{\Omega^c})_{S}=0$, which leads to $(\nabla V)_S=(\nabla V_{\Omega})_{S}$. Then, it follows from Lemma \ref{lem:3} that
\begin{equation*}
\|(\nabla V)_S\|_1=\|(\nabla V_{\Omega})_{S}\|_1\leq\|\nabla V_{\Omega}\|_1\leq\sum_{j\in\Omega} |c_{(j)}|\|\nabla h_{(j)}\|_1\leq 8\sum_{j\in\Omega}|c_{(j)}|.
\end{equation*}
Let $k=6s\log(N)$, $\|c_{\Omega}\|_1$ and $\|c_{\Omega^c}\|_1$ denote $\sum_{j\in\Omega}|c_{(j)}|$ and $\sum_{j\in\Omega^c}|c_{(j)}|$, respectively. Concerning the decay of the wavelet coefficients in Lemma \ref{lem:1}, we have $|c_{(j+1)}|\leq \tilde{C}\|\nabla V\|_1/j$. Together with the cone constraint \eqref{equ:cone3} for $\nabla V$, we have
\begin{equation*}\begin{split}
\|c_{\Omega^c}\|_1
\leq& \sum_{j=s+1}^{N^2}|c_{(j)}|\leq \tilde{C}\sum_{j=s+1}^{N^2} \frac{\|\nabla V\|_1}{j-1}\overset{(\diamond)}{\leq} C'\log\left(\frac{N^2}{s}\right)\|\nabla V\|_1\\
 \leq& C'\log\left(\frac{N^2}{s}\right) \left(2\|(\nabla V)_S\|_1+2\|\nabla \bar{X}-(\nabla \bar{X})_{s}\|_1-\frac{\alpha}{2}\|\nabla V\|_2^2+\alpha\left<\nabla X^{\rm{opt}},\nabla V\right>\right)\\
 \leq& C'\log\left(\frac{N^2}{s}\right) \left(16\|c_{\Omega}\|_1+2\|\nabla \bar{X}-(\nabla \bar{X})_{s}\|_1-\frac{\alpha}{2}\|\nabla V\|_2^2+\alpha\|\nabla X^{\rm{opt}}\|_2\|\nabla\|_2\|V\|_2\right)\\
\overset{(*)}{\leq}& C'\log\left(\frac{N^2}{s}\right) \left(16\|c_{\Omega}\|_1+2\|\nabla \bar{X}-(\nabla \bar{X})_{s}\|_1-\frac{\alpha}{2}\|\nabla V\|_2^2+\alpha\sqrt{8}\|\nabla X^{\rm{opt}}\|_2\|V\|_2\right),
\end{split}\end{equation*}
where $(\diamond)$ is due to the property of partial sum of harmonic series \cite{MR1411676}, and $(*)$ is due to the fact $\|\nabla\|_2^2\leq 8$ \cite{chambolle2004algorithm}. As we prepare to apply Proposition \ref{prop:main} to $c={\mathcal{H}} V$, we need to bound $\|\nabla V\|_2$ below in terms of $\|V\|_2=\|c\|_2$, where $\|V\|_2=\|c\|_2$ is due to Parseval's identity and the fact that $\{h_{(j)}\}$ forms an orthonormal basis for $\mathbb{C}^{N\times N}$. As $\|\nabla V\|_2 \geq \frac{1}{\sqrt{2}N}\|\nabla V\|_1$, the classical Sobolev inequality \eqref{equ:classicalsobolev} implies
\begin{equation}\label{equ:22}
\|\nabla V\|_2 \geq \frac{1}{\sqrt{2}N}\|V\|_2.
\end{equation}
Thus we have
\begin{equation}\label{equ:coneconstraint}\begin{split}
\|c_{\Omega^c}\|_1
\leq C'\log\left(\frac{N^2}{s}\right) \left(16\|c_{\Omega}\|_1+2\|\nabla \bar{X}-(\nabla \bar{X})_{s}\|_1-\frac{\alpha\|c\|_2^2}{4N^2}+\alpha\sqrt{8}\|\nabla X^{\rm{opt}}\|_2\|c\|_2\right).
\end{split}\end{equation}

       \item \emph{Tube constraint $\|{\mathcal{M}}{\mathcal{H}}^*c\|_2\leq 2\tau$.} As $\bar{X}$ and $X^{\rm{opt}}$ are in the feasible region of the model \eqref{equ:model}, for $c={\mathcal{H}} V = {\mathcal{H}} X^{\rm{opt}}-{\mathcal{H}}\bar{X}$, we have
       $$\|{\mathcal{M}}{\mathcal{H}}^*c\|_2=\|{\mathcal{M}} X^{\rm{opt}}-{\mathcal{M}}\bar{X}\|_2\leq \|{\mathcal{M}} X^{\rm{opt}}-y\|_2+\|{\mathcal{M}} \bar{X}-y\|_2\leq 2\tau.$$
\end{itemize}

Under the derived cone and tube constraints on $c$, along with the RIP condition on ${\mathcal{M}}{\mathcal{H}}^*$, Theorem \ref{thm:2} is proved by applying Proposition \ref{prop:main} and using $\gamma =16C'\log(N^2/s)\leq 32C'\log(N)$, $k = 6s\log(N)$, $\sigma=2C'\log\left(N^2/s\right)\|\nabla \bar{X}-(\nabla \bar{X})_{s}\|_1$, $E_1=\sqrt{8}\|\nabla X^{\rm{opt}}\|_2$, $E_2=\|c\|_2$, $\beta_1=\alpha C'\log\left(N^2/s\right)/(2N^2)$, and $\beta_2=\alpha C'\log\left(N^2/s\right)$. In fact, $5k\gamma^2$ with both particular $k$ and $\gamma$ leads to the required RIP order $Cs\log^3(N)$ for ${\mathcal{M}}{\mathcal{H}}^*$. Together with all these factors and Proposition \ref{prop:main}, we know that if
\begin{equation*}
\alpha \leq \frac{\sqrt{8}\sqrt{6s\log(N)}}{K_2\|\nabla X^{\rm{opt}}\|_2},
\end{equation*}
then it holds that
$$\|V\|_2=\|c\|_2\leq \sqrt{\frac{64N^2\sqrt{6s\log(N)}K_1}{\alpha K_2}\tau+\frac{8N^2}{\alpha}\|\nabla \bar{X}-(\nabla \bar{X})_{s}\|_1},$$
which leads to the estimation \eqref{equ:bound2}. \hfill $\square$

\subsection{Proof of Theorem \ref{thm:3}}\label{sec:proof3}

The proof of Theorem \ref{thm:3} follows the approach of Theorem \ref{thm:2}, in which the \emph{local coherence} of the sensing basis (Fourier) with respect to the sparsity basis (Haar wavelet) plays a major role.

\begin{definition}[Local coherence \cite{krahmer2013stable}]
The \emph{local coherence} of an orthonormal basis $\Phi=\{\phi_j\}_{j=1}^N$ of $\mathbb{C}^N$ with respect to the orthonormal basis $\Psi=\{\psi_k\}_{k=1}^N$ of $\mathbb{C}^N$ is the function $\mu^{\rm{loc}}(\Phi,\Psi)\in\mathbb{R}^N$ defined coordinate-wise by
$$\mu^{\rm{loc}}_j(\Phi,\Psi)=\sup_{1\leq k\leq N}|\left<\phi_j,\psi_k\right>|,\quad j = 1,2,\ldots,N.$$
\end{definition}

The following result indicates that, with high probabilities, signals can be stably reconstructed from subsampled measurements with the local coherence function appropriately used. It can be deemed as a finite-dimensional analog to \cite[Theorem 2.1]{rauhut2012sparse}, and a proof can be found in \cite{krahmer2013stable}.
\begin{lemma}\label{lem:fourier1}
Let $\Phi=\{\phi_j\}_{j=1}^N$ and $\Psi=\{\psi_k\}_{k=1}^N$ be two orthonormal bases of $\mathbb{C}^N$. Assume the local coherence of $\Phi$ with respect to $\Psi$ is point-wise bounded by the function $\kappa$ in the sense of
\begin{equation*}
\sup_{1\leq k \leq N}|\left<\phi_j,\psi_k\right>|\leq \kappa_j.
\end{equation*}
Fix $\delta>0$ and integers $N$, $m$, and $s$ such that $s\gtrsim\log(N)$ and $m\gtrsim \delta^{-2}\|\kappa\|_2^2s\log^3(s)\log(N)$,
and choose $m$ (possibly not distinct) indices $j\in\Omega\subset\{1,2,\ldots,N\}$ i.i.d. from the probability measure $\nu$ on $\{1,2,\ldots,N\}$ given by $v(j)=\kappa_j^2/\|\kappa\|_2^2$.

Consider the matrix $A\in\mathbb{C}^{m\times N}$ with entries $A_{j,k} = \left<\phi_j,\psi_k\right>$, $j\in\Omega,~k\in\{1,2,\ldots,N\}$,
and consider the diagonal matrix $G={\rm{diag}}(g)\in\mathbb{C}^{m\times m}$ with $g_j = \|\kappa\|_2/\kappa_j$, $j=1,\ldots,m$. Then with probability at least $1-N^{-c\log^3(s)}$, the RIC $\delta_s$ of the preconditioned matrix $\frac{1}{\sqrt{m}}GA$ satisfies $\delta_s\leq\delta$.
\end{lemma}

In particular, the following result describes the local coherence of the orthonormal Fourier basis with respect to the orthonormal Haar wavelet basis, which was initially occurred in \cite{krahmer2013stable}.
\begin{lemma}[Theorem 4 in \cite{krahmer2013stable}, slightly modified]\label{lem:fourier2}
Let $N=2^n$ be a power of 2, where $n\in\mathbb{N}^+$. The local coherence $\mu^{\rm{loc}}$ of the orthonormal two-dimensional Fourier basis $\{\varphi_{k_1,k_2}\}$ with respect to the orthonormal bivariate Haar wavelet basis $\{h^{\ell}_{j,k}\}$ in $\mathbb{C}^{N\times N} $ is bounded by
\begin{equation*}\begin{split}
\mu^{\rm{loc}}_{k_1,k_2} &\leq \kappa(k_1,k_2) :=\min\left(1,\frac{18\pi}{\max(|k_1|,|k_2|)}\right)\\
&\leq \kappa'(k_1,k_2):=\min \left(1,\frac{18\pi\sqrt2}{(|k_1|^2+|k_2|^2)^{1/2}}\right),
\end{split}\end{equation*}
and one has $\|\kappa\|_2\leq\|\kappa'\|_2\leq \sqrt{17200+502\log(N)}.$
\end{lemma}

\begin{remark}
For Theorem 4 in \cite{krahmer2013stable}, $n\geq 8$ was assumed to ensure $17200+502\log(N)\leq 2700\log(N)$ and hence $\|\kappa\|_2\leq\|\kappa'\|_2\leq 52\sqrt{\log(N)}$. We regard the assumption as a restriction on the size $N\times N$ of images, thus we remove this assumption and adopt the bound $\sqrt{17200+502\log{N}}$ in our following proof. Besides, it was conjectured in \cite{krahmer2013stable} that the factor 2700 is due to lack of smoothness for the Haar wavelets, and this factor might be removed by considering smoother wavelets.
\end{remark}

\noindent \emph{Proof of Theorem \ref{thm:3}.} Let $P\in\mathbb{C}^{m\times m}$ be the diagonal matrix encoding the weights in the noise model. That is, $P={\rm{diag}}(\rho)$, where, for $\kappa'$ as in Lemma \ref{lem:fourier2}, $\rho\in\mathbb{C}^m$ is a vector converted from the matrix
\begin{equation*}
\rho(k_1,k_2)=\frac{\|\kappa'\|_2}{\kappa'(k_1,k_2)}=C\sqrt{1+\log(N)}\max\left(1,\frac{(|k_1|^2+|k_2|^2)^{1/2}}{18\pi}\right),\quad (k_1,k_2)\in\Omega.
\end{equation*}
Note that $Pg = \rho\circ g$ for $g\in\mathbb{C}^{m}$. Together with the particular incoherence estimate in Lemma \ref{lem:fourier2}, Lemma \ref{lem:fourier1} implies that with probability at least $1-N^{-2c\log^3(s)}$ (as $c$ is a generic constant, the factor $2$ of $c$ is removed in the statement of Theorem \ref{thm:3}), $\mathcal{A}:=\frac{1}{\sqrt{m}}P{\mathcal{F}}_{\Omega}{\mathcal{H}}^*$ has the RIP of order $s$ and level $\delta<0.6$ once $s\gtrsim \log(N^2)\gtrsim\log(N)$ and
\begin{equation*}
m\gtrsim s\delta^{-2}\log^3(s)\log^2(N^2)\gtrsim s\delta^{-2}\log^3(s)\log^2(N).
\end{equation*}
By the assumption $m\gtrsim s\log^3(s)\log^5(N)$ (in fact, we shall assume $m\gtrsim s\delta^{-2}\log^3(s)\log^5(N)$), we can assume that $\mathcal{A}$ has the RIP of order $\bar{s}= Cs\log^3(N)$ and level $\delta<0.6$, where $C$ is the constant derived in Theorem \ref{thm:2}.
Moreover, let $V=X^{\rm{opt}}-\bar{X}$ and apply Proposition \ref{prop:main} again to $c={\mathcal{H}} V$, where $c_{(1)}:=c_{(1)}(V)$ denotes the Haar coefficient corresponding to the constant wavelet, and $c_{(j)}:=c_{(j)}(V)$ ($j\geq2$) denotes the $(j-1)$-st largest-magnitude Haar coefficient among the remaining. To apply Proposition \ref{prop:main}, we need to find cone and tube constraints for $c={\mathcal{H}} V$.
\begin{itemize}
       \item \emph{Cone constraint on $c={\mathcal{H}} V$,} which is the same as \eqref{equ:coneconstraint} in the proof of Theorem \ref{thm:2}.
       \item \emph{Tube constraint $\|\mathcal{A} c\|_2=\|\mathcal{A}{\mathcal{H}} V\|_2\leq \sqrt{2}\tau$,} since
       \begin{equation*}\begin{split}
       m\|\mathcal{A}{\mathcal{H}} V\|_2^2 &=\|P{\mathcal{F}}_{\Omega}{\mathcal{H}}^*{\mathcal{H}} V\|_2^2=\|\rho\circ({\mathcal{F}}_{\Omega} V)\|_2^2\\
       &\leq  \|\rho\circ({\mathcal{F}}_{\Omega} X^{\rm{opt}}-b)\|_2^2+ \|\rho\circ({\mathcal{F}}_{\Omega} \bar{X}-b)\|_2^2\leq2m\tau^2.
       \end{split}\end{equation*}
\end{itemize}

The rest is similar to the proof of Theorem \ref{thm:2}, and the only trivial difference is the tube constraint, where $2\tau$ there is replaced by $\sqrt{2}\tau$ here. Hence, we omit the following steps, and the estimation for the setting in this theorem, with constants removed, is the same as \eqref{equ:bound2}.  \hfill $\square$

\section{Numerical experiments}\label{sec:numerical}

We now report some experimental results to validate the quality of reconstruction and numerical solvability of the enhanced TV model \eqref{equ:model}. As mentioned, the model \eqref{equ:model} is of difference-of-convex, and it can be solved by some well-developed algorithms in the literature. We include the details of an algorithm in Appendix \ref{sec:algorithm}. For comparison, we consider the TV model \eqref{equ:TVmodel} and the $\text{TV}_a-\text{TV}_i$ model in \cite{lou2015weighted}. In our experiments, the TV model \eqref{equ:TVmodel} is solved by the split Bregman method studied in \cite{goldstein2009split}, and the $\text{TV}_a-\text{TV}_i$ model is solved by the difference-of-convex functions algorithm (DCA) with subproblems solved by the split Bregman method in \cite{lou2015weighted}. Details of tuned parameters of these algorithms are stated in Appendix \ref{sec:algorithm}. As displayed in Figure \ref{fig:test}, we test the standard \emph{Shepp--Logan phantom}, three more synthetic piecewise-constant images (\emph{Shape}, \emph{Circle}, and \emph{USC Mosaic}), two natural images (\emph{Pepper} and \emph{Clock}), and two medical images (\emph{Spine} and \emph{Brain}). Two sampling strategies are considered in our experiments. The first one is the \emph{radial lines} sampling, and the other one is the strategy \eqref{equ:sampling} proposed in Theorem \ref{thm:3}, which is referred to as the \emph{MRI-desired} sampling strategy below. All codes were written by MATLAB R2021b, and all numerical experiments were conducted on a laptop (16 GB RAM, Intel CoreTM i7-9750H Processor) with macOS Monterey 12.1.
\begin{figure}[htbp]
  \centering
  \includegraphics[width=12.5cm]{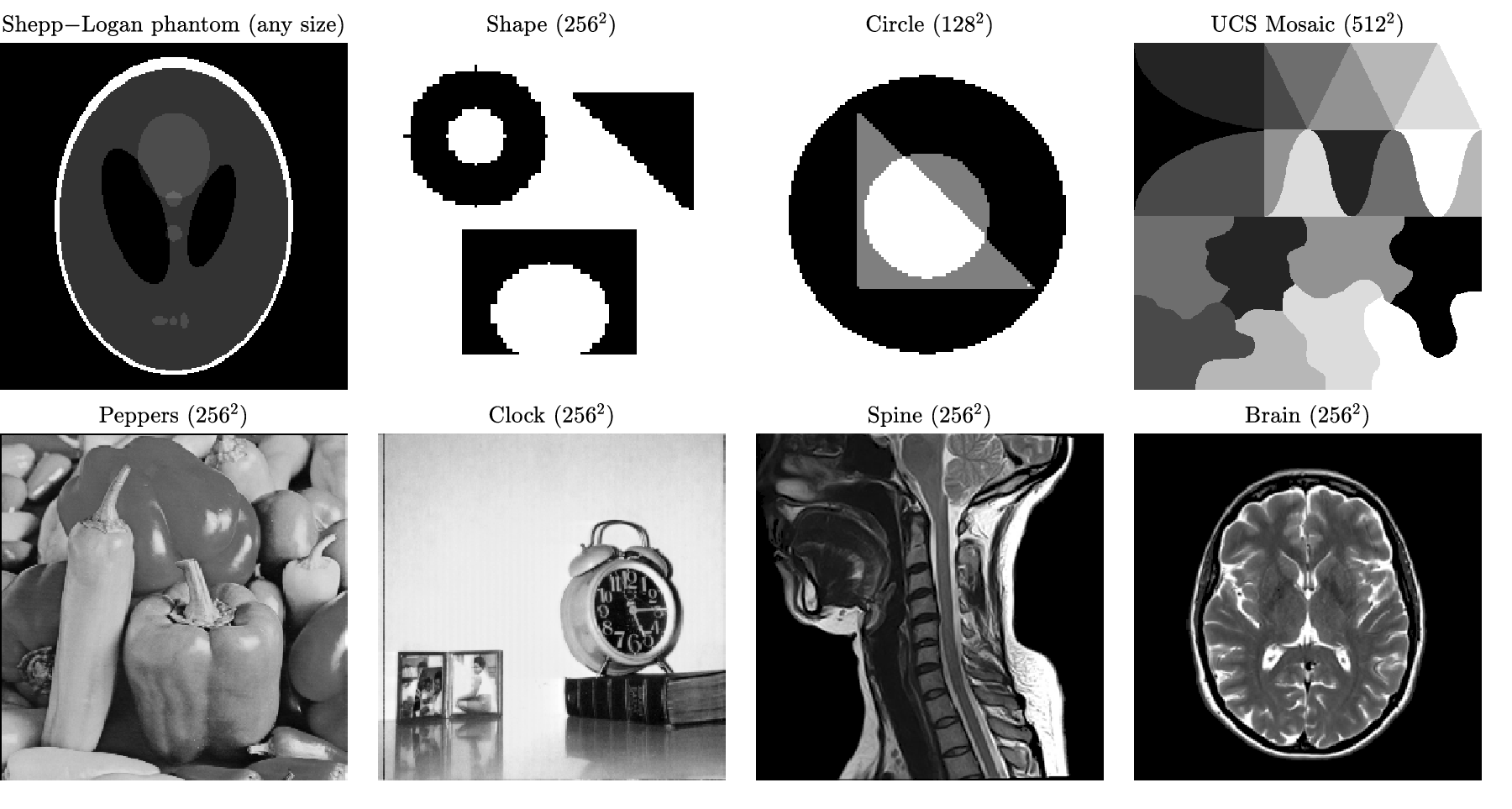}
  \caption{Test images.}\label{fig:test}
\end{figure}

\begin{figure}[htbp]
  \centering
  \includegraphics[width=\textwidth]{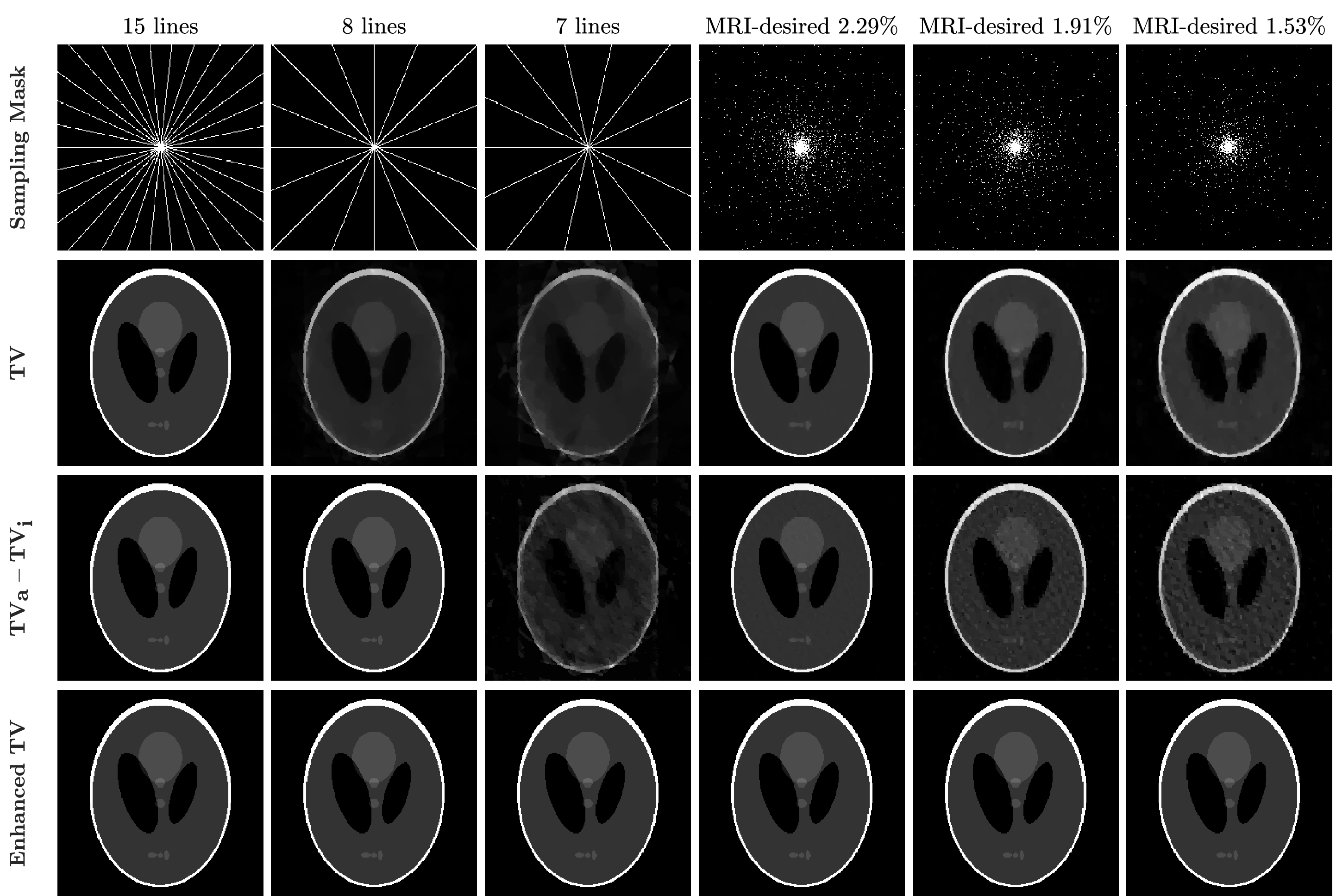}
  \caption{Shepp--Logan phantom: Comparison of three models with radial line-sampled and MRI-desired measurements.}\label{fig:phantom}
\end{figure}

\noindent \textbf{Example \#1: Shepp--Logan phantom.} The Shepp--Logan phantom is standard in the image reconstruction literature. Experiments for this image are organized into three parts. The first part concentrates on the reconstruction of the Shepp--Logan phantom of size $256\times 256$ from noise-free measurements, and $\alpha$ is fixed as $0.8$ in the enhanced TV model \eqref{equ:model}. We sample along 15, 8, and 7 radial lines, corresponding to sampling rates 6.44\%, 3.98\%, and 3.03\%, respectively, and take MRI-desired measurements with rates 2.29\%, 1.91\%, and 1.53\%. The results shown in Figure \ref{fig:phantom} suggest that the enhanced TV model \eqref{equ:model} produces reconstruction with good accuracy in all six sampling settings, and reconstruction quality is much better than those in comparison when the amount of samples is limited (e.g., 7 radial lines and 1.53\% MRI-desired measurements). This observation verifies the result in Section \ref{sec:discussion}. That is, when $\tau=0$, the reconstruction error bound \eqref{equ:bound3} for the enhanced TV model \eqref{equ:model} is tighter than \eqref{equ:TVrecoveryimproved} for the TV model \eqref{equ:model} with a limited amount of measurements. As mentioned in Section \ref{sec:discussion}, such a result also pertains to the comparison between the enhanced TV model \eqref{equ:model} and the $\text{TV}_a-\text{TV}_i$ model in \cite{lou2015weighted}.

For comparison, we also report relative errors in the Frobenius sense and SSIM values in Table \ref{tab:error}. Advantages of the enhanced TV model \eqref{equ:model} are shown when the available measurements are limited (e.g., when the sampling rate is below 3.03\%). When the measurements are relatively sufficient, e.g., in the cases of 15 lines and eight lines, the enhanced TV model \eqref{equ:model} does not produce reconstruction with the least error. We see that the outperformance of the enhanced TV model \eqref{equ:model} is not maintained when the measurements become sufficient, while the difference is too tiny to be visually observed. Besides, it is worth noting that SSIM values in all six sampling settings are 1.0000 for the enhanced TV model \eqref{equ:model}, and the stability of this model with respect to the amounts of measurements is well illustrated for the Shepp--Logan phantom images.

\begin{table}[htbp]
  \centering
  \footnotesize
  \setlength{\abovecaptionskip}{0pt}
\setlength{\belowcaptionskip}{10pt}
  \caption{Relative errors and SSIM value of the reconstructed images in Figure \ref{fig:phantom}.}\label{tab:error}
  \begin{tabular}{c|c|c|c}\hline
    & TV & $\text{TV}_a-\text{TV}_i$ &  Enhanced TV \\ \hline
15 lines (6.44\%) & 1.924E-13 (1.0000) & \textbf{7.845E-14} (1.0000) & 2.977E-12 (1.0000) \\\hline
8 lines (3.98\%) & 0.2456 (0.6764) & \textbf{3.852E-09} (1.0000) & 7.841E-07 (1.0000) \\\hline
7 lines (3.03\%) & 0.4819 (0.4612) & 0.3968 (0.5209) & \textbf{1.608E-06} (1.0000)  \\\hline
MRI-desired (2.29\%) & 0.0415 (0.9890) & 0.0266 (0.9896) & \textbf{8.069E-06} (1.0000)\\\hline
MRI-desired (1.91\%) & 0.1575 (0.8937) & 0.1837 (0.8404) & \textbf{2.324E-05} (1.0000) \\\hline
MRI-desired (1.53\%) & 0.2826 (0.7473) & 0.2983 (0.7374) & \textbf{8.456E-05} (1.0000)  \\\hline
  \end{tabular}
\end{table}

The second part illustrates the robustness of the enhanced TV model \eqref{equ:model} with respect to noise. We still fix $\alpha$ as $0.8$ in the model \eqref{equ:model}, and we take measurements along 15 lines (corresponding to 6.44\% sampling rate) and use 6.5\% MRI-desired samples. The Fourier measurements are perturbed by Gaussian noise with standard derivations (``\texttt{std}" for short) of 0.04, 0.06, and 0.08, respectively. The contamination process is implemented in MATLAB commands: For any image $X$ with size $N\times N$, we first compute its Fourier measurements by the fast Fourier transform (FFT), i.e., \texttt{F=fft2(X)/N}. Then we perturb \texttt{F} by \texttt{F=F+1/sqrt(2)*(std*randn(size(F))+std*1i*randn(size(F)))}. Relative errors and SSIM values listed in Table \ref{tab:errornoisy} show that the enhanced TV model \eqref{equ:model} is the most robust one. In particular, in terms of the SSIM values, the enhanced TV model \eqref{equ:model} produces much better reconstruction quality, and the superiority is more apparent when the level of noise increases. These results assert the theoretical result in Section \ref{sec:discussion} that the enhanced TV model \eqref{equ:model} has a tighter reconstruction error bound than the TV model \eqref{equ:TVmodel} and the $\text{TV}_a-\text{TV}_i$ model in \cite{lou2015weighted} when the level of noise is relatively large.

\begin{table}[htbp]
  \centering
    \footnotesize
\setlength{\abovecaptionskip}{0pt}
\setlength{\belowcaptionskip}{10pt}
  \caption{Relative errors and SSIM values of the reconstructed images in Figure \ref{fig:phantom}, with three levels of noise std = 0.04, 0.06, and 0.08.}\label{tab:errornoisy}
  \begin{tabular}{c|c|c|c}\hline
 & TV & $\text{TV}_a-\text{TV}_i$ & Enhanced TV \\ \hline
15 lines (6.44\%), std = 0.04 & 0.1796 (0.5759) & 0.1860 (0.4534) & \textbf{0.0921} (0.9531)  \\\hline
15 lines (6.44\%), std = 0.06 & 0.2506 (0.4866) & 0.2748 (0.3161) & \textbf{0.1038} (0.9490)  \\\hline
15 lines (6.44\%), std = 0.08 & 0.3111 (0.4265) & 0.3535 (0.2448) & \textbf{0.1496} (0.9359)  \\\hline
MRI-deisred (6.50\%), std = 0.04 & 0.1041 (0.7322) & 0.1376 (0.5721) & \textbf{0.0873} (0.9588) \\\hline
MRI-deisred (6.50\%), std = 0.06 & 0.1498 (0.6101) & 0.2082 (0.4179) & \textbf{0.1393} (0.9477) \\\hline
MRI-deisred (6.50\%), std = 0.08 & 0.1914 (0.5213) & 0.2764 (0.3243) & \textbf{0.1674} (0.9396) \\\hline
  \end{tabular}
\end{table}

\begin{figure}[htbp]
  \centering
  \includegraphics[width=13cm]{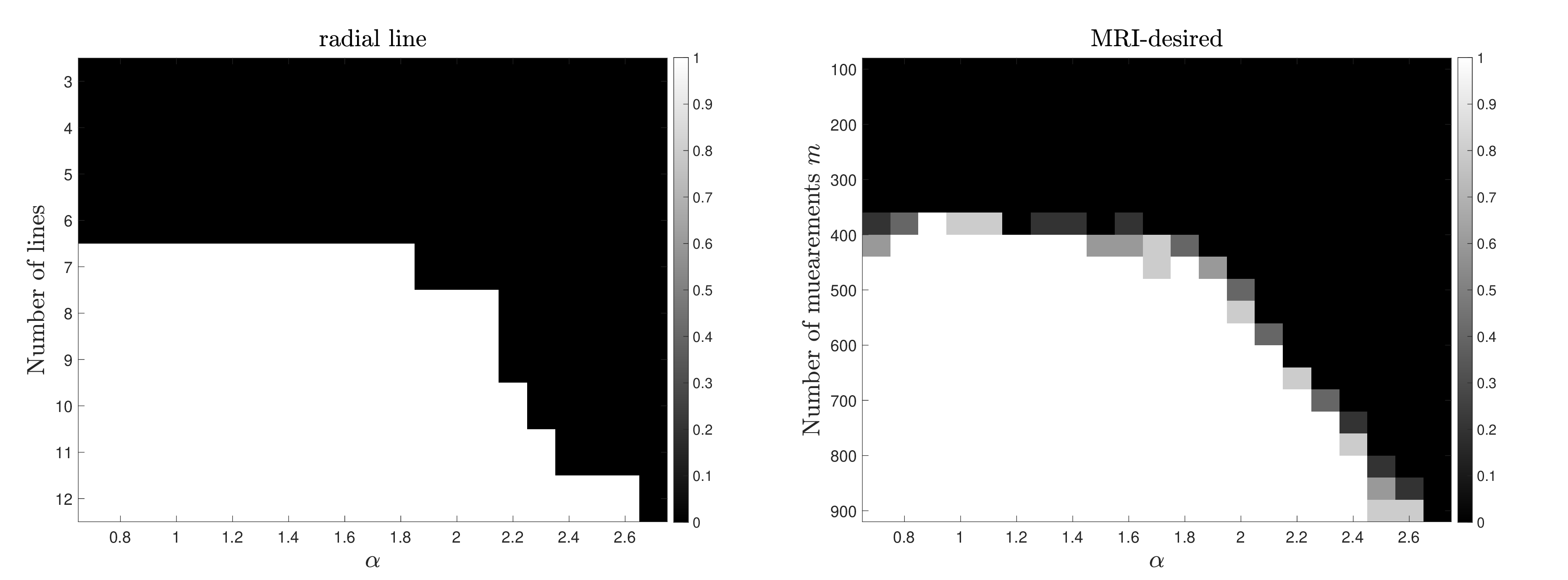}
  \caption{Phase transitions with respect to $m$ and $\alpha$.}\label{fig:phase}
\end{figure}

The third part is focused on the phase transition of the success rates of reconstruction. A reconstruction is recognized as \emph{successful} if the relative error of the reconstructed image is less than $10^{-3}$. We consider the Shepp--Logan phantom with size $64\times 64$ in this part. We choose $\alpha$ among $\{0.7,0.8,\ldots,2.7\}$ for the enhanced TV model \eqref{equ:model}, and choose the number of measurements $m$ from 3 to 12 radial lines for radial sampling, and among $\{100,140,180,\ldots,900\}$ for MRI-desired sampling. For each case, we test five times and report the success rate. According to Theorem \ref{thm:3}, stable reconstruction can be achieved if samples are enough in the sense of \eqref{equ:m} and the model parameter $\alpha$ is bounded in the sense of \eqref{equ:alpha3}. The results in Figure \ref{fig:phase} assert that a successful reconstruction via the enhanced TV model \eqref{equ:model} requires relatively sufficient samples and a reasonably bounded parameter $\alpha$, thus validating results in Theorem \ref{thm:3}.

\begin{figure}[htbp]
  \centering
  \includegraphics[width=13.5cm]{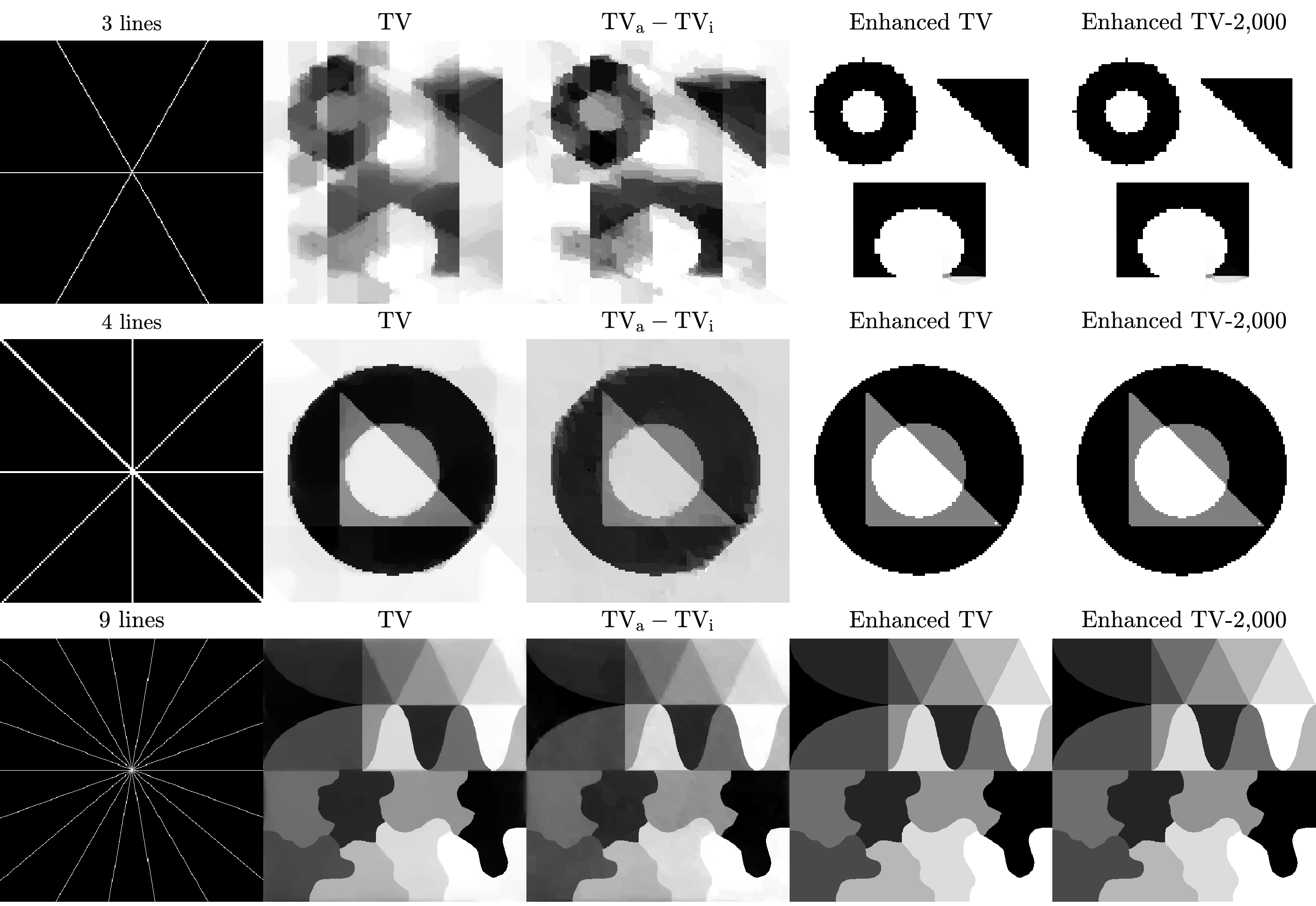}
  \caption{Shape, Circle, and USC Mosaic: Comparison of three models with limited measurements.}\label{fig:otherimages}
\end{figure}

\noindent \textbf{Example \#2: Synthetic images.} Example \#1 shows the superiority of the enhanced TV model \eqref{equ:model} for Shepp--Logan phantom with limited samples, and one purpose of the following study is to further assert this superiority. We consider the radial line sampling and validate this superiority by testing three synthetic images: Shape, Circle, and USC Mosaic. We also fix $\alpha = 0.8$ in the enhanced TV model \eqref{equ:model}. When the number of measurements is limited enough, all three models cannot generate good reconstruction. Bearing in mind that the criteria of the limitation on the amount of measurements are different for three models, we now show some cases that the reconstruction via the enhanced TV model \eqref{equ:model} is particularly good while those via the TV model \eqref{equ:TVmodel} and the $\text{TV}_a-\text{TV}_i$ model in \cite{lou2015weighted} may fail. Reconstruction results are displayed in Figure \ref{fig:otherimages}, and relative errors and SSIM values are reported in Table \ref{tab:otherimages}. From both Figure \ref{fig:otherimages} and Table \ref{tab:otherimages}, the reconstruction of the enhanced TV model \eqref{equ:model} is significantly better than the other two models.

We also take this example to test how the inner iterations can affect the overall performance of the algorithms under comparison. The algorithm presented in Appendix \ref{sec:algorithm} adopts DCA as the outer iteration and uses the ADMM to solve each DCA subproblem. When the maximum number of inner ADMM iterations is increased from 1,000 to 2,000, the numerical results are reported in the fifth column of Figure \ref{fig:otherimages}, labeled as ``Enhanced TV-2,000''. We see that even if the enhanced TV model \eqref{equ:model} with at most 1,000 inner iterations is good enough to generate a satisfactory reconstruction, e.g., for Circle and USC Mosaic, more inner iterations can further reduce the relative errors by up to several orders of magnitude. This observation provides a simple recipe for higher-accuracy reconstruction.

\begin{table}[htbp]
  \centering
    \footnotesize
\setlength{\abovecaptionskip}{0pt}
\setlength{\belowcaptionskip}{10pt}
  \caption{Relative errors and SSIM values of the reconstructed images in Figure \ref{fig:otherimages}.}\label{tab:otherimages}
  \begin{tabular}{c|c|c|c||c}\hline
 & TV & $\text{TV}_a-\text{TV}_i$ & Enhanced TV &  Enhanced TV-2,000 \\ \hline
Shape (1.29\%) & 0.3094 (0.5466) & 0.2503 (0.5458) & \textbf{0.0266} (0.9932) & 0.0261 (0.9937) \\\hline
Circle (3.86\%) & 0.0394 (0.9705) & 0.0498 (0.9430) & \textbf{7.411E-08} (1.0000) & 6.815E-13 (1.0000) \\\hline
USC Mosaic (1.95\%) & 0.0405 (0.9032) & 0.0439 (0.9024) & \textbf{8.013E-05} (1.0000) & 4.206E-07 (1.0000) \\\hline
  \end{tabular}
\end{table}

\begin{figure}[htbp]
  \centering
  \includegraphics[width=13cm]{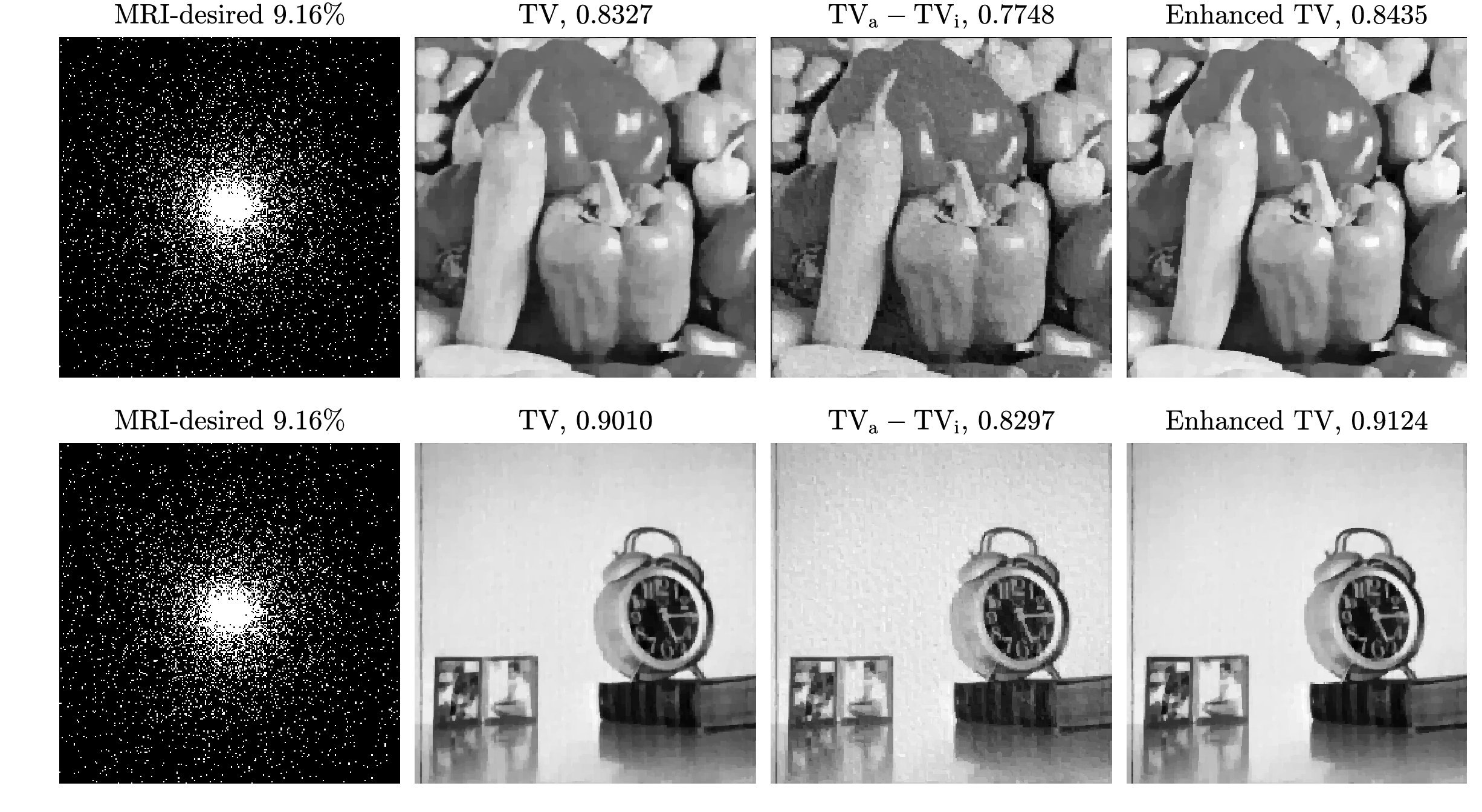}
  \caption{Peppers and Clock: Comparison of three models with the MRI-desired sampling. SSIM values are also reported in the titles of each reconstruction.}\label{fig:naturalimages}
\end{figure}

\noindent \textbf{Example \#3: Natural images.} We then test two natural images: Peppers and Clock. We fix $\alpha$ as $1$ in the enhanced TV model \eqref{equ:model}. In Figure \ref{fig:naturalimages}, we display the reconstruction of both images from 9.16\% MRI-desired samples. Furthermore, we report relative errors in the Frobenius sense and SSIM values of each reconstruction in Table \ref{tab:error2}, from MRI-desired samples of rates 9.16\%, 13.7\%, 18.3\%, and 22.9\%. The superiority of the enhanced TV model \eqref{equ:model} is further validated.

It is worth noting that the enhanced TV model \eqref{equ:model} performs less effectively for reconstructing natural images than images in Examples \#1 and \#2 because these natural images have more complicated (non-piecewise-constant) edges. It is not surprising that the enhanced TV model \eqref{equ:model} is less effective for these images because, nevertheless, it is a generalization of the TV model \eqref{equ:TVmodel}. Thus it keeps the main feature of the TV regularization for recovering piecewise-constant images while it can additionally reduce the loss of contrast.

\begin{table}[htbp]
  \centering
     \footnotesize
  \setlength{\abovecaptionskip}{0pt}
\setlength{\belowcaptionskip}{10pt}{}
  \caption{Relative errors and SSIM values of reconstructions of two natural images with various sampling rates.}\label{tab:error2}
  \begin{tabular}{c|c|c|c}\hline
    & TV & $\text{TV}_a-\text{TV}_i$ & Enhanced TV \\ \hline
Peppers (9.16\%)  & 0.0771 (0.8327) & 0.0823 (0.7748) & \textbf{0.0718} (0.8435) \\\hline
Peppers (13.73\%) & 0.0597 (0.8793) & 0.0624 (0.8409) & \textbf{0.0536} (0.8908) \\\hline
Peppers (18.31\%) & 0.0447 (0.9139) & 0.0498 (0.8800) & \textbf{0.0414} (0.9208) \\\hline
Peppers (22.89\%) & 0.0388 (0.9292) & 0.0424 (0.9035) & \textbf{0.0351} (0.9358) \\\hline
Clock (9.16\%)    & 0.0404 (0.9010) & 0.0440 (0.8297) & \textbf{0.0379} (0.9124) \\\hline
Clock (13.73\%)   & 0.0288 (0.9356) & 0.0319 (0.8884) & \textbf{0.0272} (0.9421) \\\hline
Clock (18.31\%)   & 0.0213 (0.9563) & 0.0246 (0.9218) & \textbf{0.0203} (0.9592) \\\hline
Clock (22.89\%)   & 0.0182 (0.9647) & 0.0205 (0.9393) & \textbf{0.0169} (0.9674) \\\hline

  \end{tabular}
\end{table}

\begin{figure}[htbp]
  \centering
  \includegraphics[width=13cm]{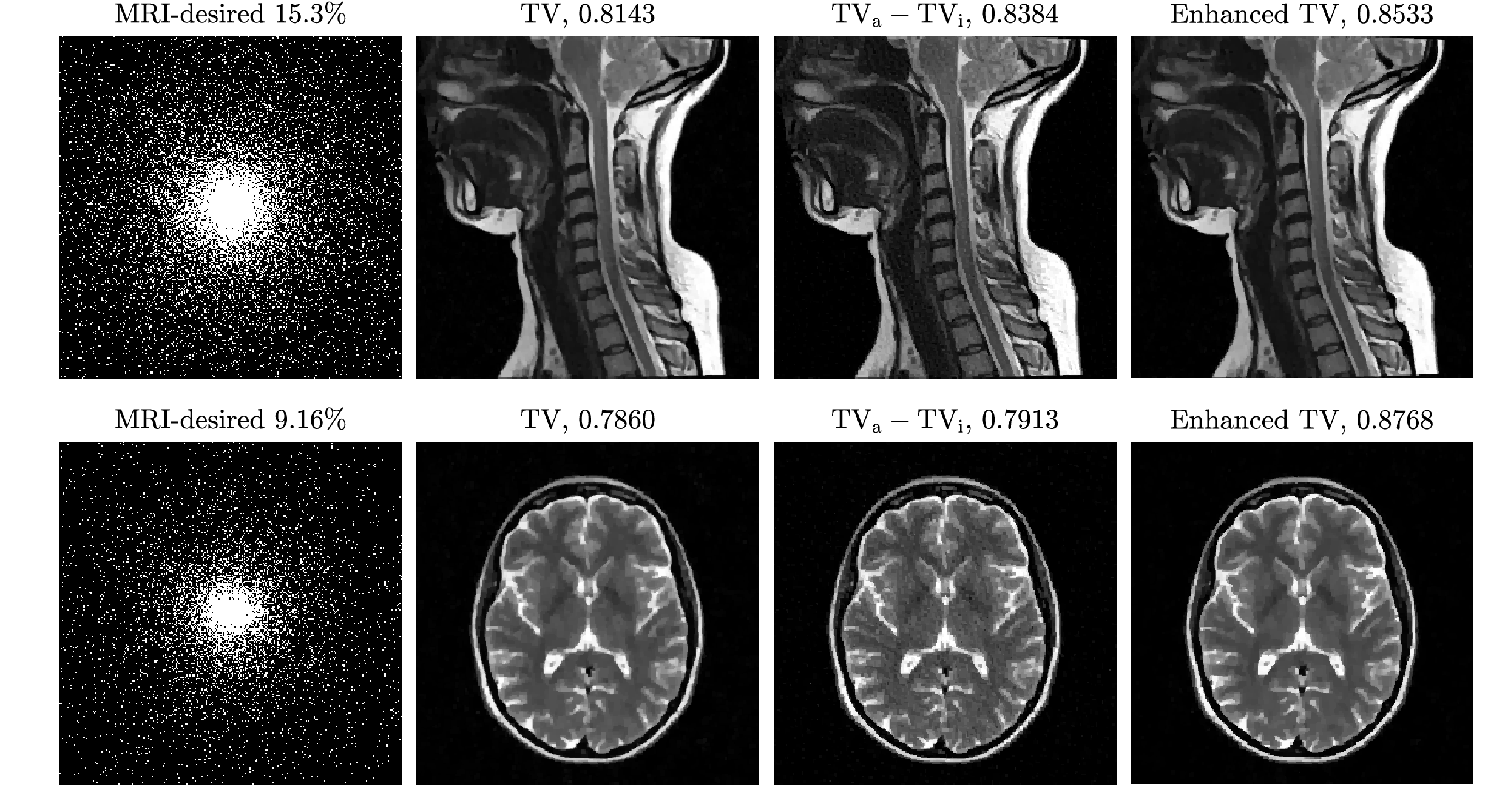}
  \caption{Spine and Brain: Comparison of three models on medical images with the MRI-desired sampling. SSIM values are also reported in the titles of each reconstruction.}\label{fig:mriimages}
\end{figure}

\begin{figure}[htbp]
  \centering
  \includegraphics[width=13cm]{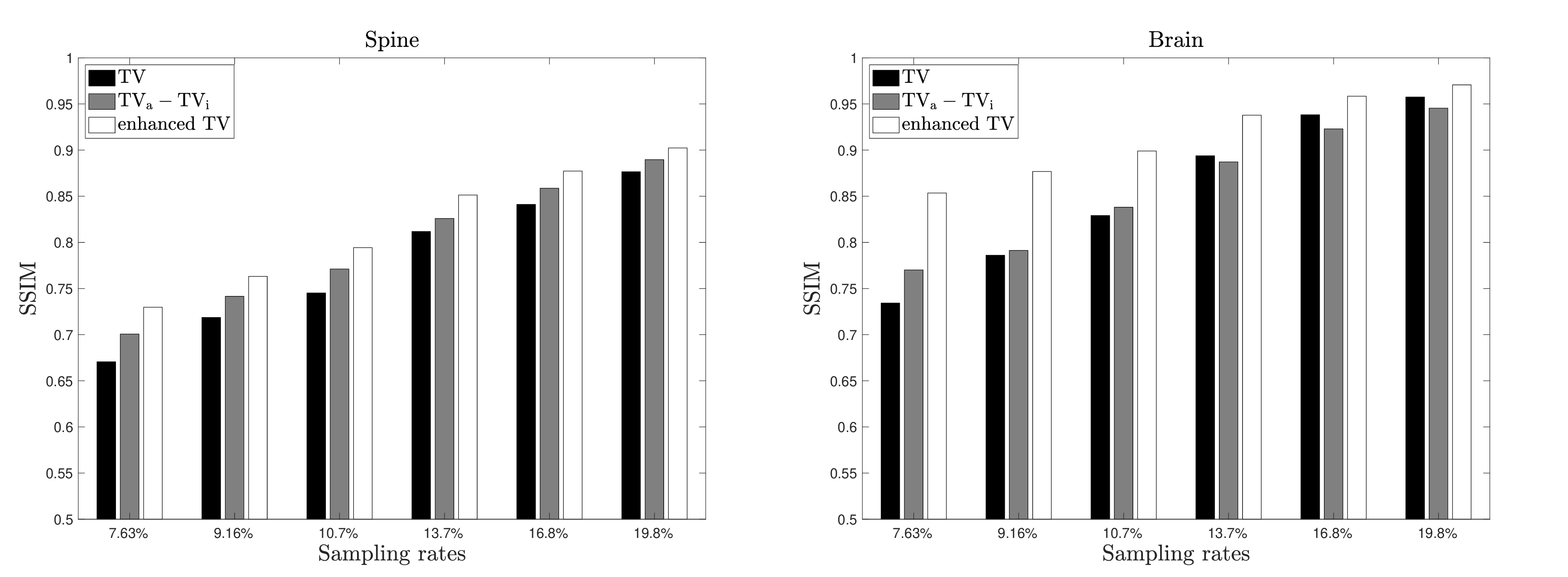}
  \caption{Spine and Brain: SSIM values of reconstructions with various sampling rates.}\label{fig:percentage}
\end{figure}

\noindent \textbf{Example \#4: Medical images.} Finally, we test two medical images: Spine and Brain. We again fix $\alpha$ as $1$ in the enhanced TV model \eqref{equ:model}. Moreover, we take 15.3\% MRI-desired samples for the reconstruction of Spine and 9.16\% for Brain, and the reconstructed images are displayed in Figure \ref{fig:mriimages}. It is shown that the enhanced TV model \eqref{equ:model} produces better reconstructions than the other models. We test more sampling rates and report the SSIM values of reconstructions with each rate in Figure \ref{fig:percentage}. It is easy to see that the superiority of the enhanced TV model \eqref{equ:model} is more apparent when the sampling rate is relatively low. Thus, the enhanced TV model \eqref{equ:model} is preferred when measurements are limited. Similar to Example \#3, the enhanced TV model \eqref{equ:model} performs less effectively for Example \#4 than Examples \#1 and \#2 due to the non-piecewise-constant edges of these medical images.

\section{Conclusions}\label{sec:conclusion}
We focused on enhancing the canonical constrained total variational (TV) minimization model for image reconstruction by the spingback regularization in our previous work \cite{an2021springback}. The enhanced TV model improves the original TV model with an additional backward diffusion process so that the loss of contrast can be further reduced. We theoretically established the reconstruction guarantees using the enhanced TV model \eqref{equ:model} for non-adaptive subsampled linear RIP measurements and variable-density subsampled Fourier measurements, respectively. For non-adaptive linear RIP measurements, the requirement on the RIP level $\delta$ was relaxed from $\delta<1/3$ (derived for the TV model \eqref{equ:TVmodel}; see \cite{needell2013stable}) to $\delta<0.6$. The reconstruction error bounds estimated in Theorems \ref{thm:1} and \ref{thm:2} suggest reasonable reconstruction error estimations for the TV model \eqref{equ:TVmodel} when $\delta\rightarrow 0.6$, in which case the bounds derived in \cite{needell2013stable} for the TV model \eqref{equ:TVmodel} tend to be infinity. For variable-density sampled Fourier measurements, the required least amount of measurements of the enhanced TV model \eqref{equ:model} was shown to be around 30.86\% of that established in \cite{krahmer2013stable} for the TV model \eqref{equ:TVmodel}. This improvement is due to the relaxation of the requirement on $\delta$. 

Recall that we only consider the anisotropic TV, and proofs of the main theoretical results can be easily generalized to the isotropic TV case. In addition, our results can be generalized from several other perspectives. For example, one can consider other sampling strategies, such as those in \cite{adcock2021improved,poon2015role} for Fourier samples as considered in Theorem \ref{thm:3}. For the guarantees analysis with Fourier measurements, noise is measured by the weighted $\ell_2$-norm (see \eqref{equ:fouriermodel}), and then one can consider some other norms to measure noise such as those in \cite{adcock2021improved,poon2015role}. Our theoretical results for two-dimensional images can also be extended to higher dimensional signals, as considered in \cite{adcock2021improved,needell2013near}. It seems also promising to consider applying the enhanced TV model \eqref{equ:model} to other problems such as image inpainting and super-resolution problems, combining the enhanced TV regularization \eqref{equ:penalty} with other data fidelity terms to model some problems such as image segmentation and motion estimation, and using the enhanced TV regularization \eqref{equ:penalty} in combination with other widely-used convex and/or non-convex regularizers to model various more challenging image processing problems.

\appendix

\section{The enhanced TV model \eqref{equ:model} in a continuum setting}\label{sec:continuum}
Let $u:\Omega\rightarrow \mathbb{R}$ be an image, where the image domain $\Omega$ is a bounded and open subset of $\mathbb{R}^2$. The TV denoising model in \cite{rudin1992nonlinear} for a noisy image $u_0:\Omega\rightarrow \mathbb{R}$ is formulated as
\begin{equation}\label{equ:TVdenosing}
\min_u~\mathcal{E}_{\rm{TV}}(u):= \int_{\Omega}|\nabla u| \textrm{d}x +\frac{\mu}{2}\int_{\Omega}(u(x)-u_0(x))^2\textrm{d}x,
\end{equation}
where $x=(x_1,x_2)\in\Omega$, $|\nabla u|=\sqrt{(\partial_{x_1} u)^2+(\partial_{x_2} u)^2}$, and $\mu>0$ balances the TV term and the data fidelity term. Note that the isotropic TV proposed in \cite{rudin1992nonlinear} is used in the model \eqref{equ:TVdenosing}. Though the anisotropic TV defined in \cite{esedoglu2004decomposition} is used in the enhanced TV regularization \eqref{equ:penalty}, the main purpose of this appendix is to explain how the TV is enhanced in the sense of \eqref{equ:penalty}. Thus, we adopt the model \eqref{equ:TVdenosing} for simplicity. We refer the reader to \cite{MR2139257} for the anisotropic TV flow. More specifically, the enhanced (isotropic) TV denoising model in a continuum setting can be written as
\begin{equation}\label{equ:minfunctional}
\min_u~\mathcal{E}_{\rm{ETV}}(u):= \int_{\Omega}|\nabla u| \textrm{d}x-\frac{\alpha}{2}\int_{\Omega}|\nabla u|^2\textrm{d}x +\frac{\mu}{2}\int_{\Omega}(u(x)-u_0(x))^2\textrm{d}x.
\end{equation}
Then, by computing the first-order variation of the functional, the Euler--Lagrange equation associated with the energy functional $\mathcal{E}_{\rm{ETV}}(u)$ in the distributional sense is
\begin{equation}\label{equ:EL}
0 =-\nabla \cdot\left[\frac{\nabla u}{|\nabla u|}\right]+\alpha\Delta u +\mu (u-u_0)\quad \text{with}\quad \frac{\partial u}{\partial \textbf{n}}\bigg|_{\partial \Omega}=0,
\end{equation}
where $\textbf{n}$ denotes the outer normal derivative along the boundary $\partial \Omega$ of $\Omega$.

Alternatively, as \cite{rudin1992nonlinear}, we could use the \emph{gradient descent marching with artificial time} $t$. That is, the solution procedure of the Euler--Lagrange equation \eqref{equ:EL} uses a parabolic equation with time $t$ as an evolution parameter. This means, for $u:\Omega\times [0,T]\rightarrow\mathbb{R}$, we solve
\begin{equation}\label{equ:backwarddiffusion}
u_t=-\frac{\partial \mathcal{E}_{\rm{ETV}}}{\partial u}= \nabla \cdot\left[\frac{\nabla u}{|\nabla u|}\right]-\alpha\Delta u -\mu (u-u_0) \quad\text{for }t>0,~x\in\Omega,
\end{equation}
with a given initial condition $u(x,0)$ and the boundary condition $\frac{\partial u}{\partial \textbf{n}}|_{\partial \Omega}=0$. Note that there is a backward diffusion term $-\alpha\Delta u$ in the evolution equation \eqref{equ:backwarddiffusion}. Thus, as $t$ increases, we approach a denoised and deblurred version of the image if the blur is assumed to follow such a diffusion process.

If the energy functional $\mathcal{E}_{\rm{ETV}}(u)$ has a minimum, then the minimizer must satisfy the Euler--Lagrange equation \eqref{equ:backwarddiffusion}. Certainly, the existence of the minimizer of $\mathcal{E}_{\rm{ETV}}$ is unknown for an arbitrary $\alpha$. On the other hand, with $\alpha < \mu \inf_{x\in\Omega}\frac{|u(x)|^2}{|\nabla u(x)|^2}$, the Lagrangian
\begin{equation*}
\mathcal{L}_{\rm{ETV}}(\nabla u, u, x) :=|\nabla u| -\frac{\alpha}{2}|\nabla u|^2+\frac{\mu}{2}(u(x)-u_0(x))^2
\end{equation*}
is bounded below by $|\nabla u(x)| + \frac{\mu-\alpha}{2}| u(x)|^2-\mu u(x)u_0(x)+|u_0(x)|^2$, which is a convex function with respect to variables $\nabla u$ and $u$. Hence, $\mathcal{E}_{\rm{ETV}}$ is bounded below, and any stationary point $u^*$ of $\mathcal{E}_{\rm{ETV}}$ (including global and local minimizers) must be finite and satisfy the corresponding Euler--Lagrange equation \eqref{equ:backwarddiffusion} involving the backward diffusion term. This requirement on $\alpha$ explains the rationale of the assumption on the upper bound of $\alpha$ in Theorems \ref{thm:1}, \ref{thm:2}, and \ref{thm:3} (e.g., $\alpha \leq \frac{\sqrt{48s\log(N)}}{K_2\|\nabla X^{\rm{opt}}\|_2}$ in Theorems \ref{thm:2} and \ref{thm:3}).

\section{Implementation details for reproducing Figure \ref{fig:motivation}}\label{sec:implementation}
For denoising, let the noisy image be $y\in\mathbb{C}^{N\times N}$ be $y={\bar{X}}+e$. The denoising model using the enhanced TV regularization (\ref{equ:penalty}) is formulated as
\begin{equation}\label{equ:denoising}
\min_{X\in\mathbb{C}^{N\times N}}~\|\nabla X\|_1-\frac{\alpha}{2}\|\nabla X\|_2^2+\frac{\mu}{2}\|y-X\|_2^2,
\end{equation}
where $\mu>0$ is a parameter balancing the enhanced TV regularization term and the data fidelity term. Note that the model \eqref{equ:denoising} is the discretization of the model \eqref{equ:minfunctional}. The model \eqref{equ:denoising} can be solved by the DCA in \cite{tao1997convex,tao1998dc}, and its subproblems can be solved by the splitting Bregman iteration in \cite{goldstein2009split}. We summarize the resulting algorithm as Algorithm \ref{alg:DCA2} below, in which MaxDCA denotes the maximum number of the DCA iterations and MaxBreg denotes is the maximum number of the Bregman iterations.
\begin{algorithm}[htbp]\label{alg:DCA2}
\caption{Solving the unconstrained denoising model \eqref{equ:denoising}}
\KwIn{Define $X^0 = 0$, $z=0$, $k=0$, $d_x=d_y=0$, MaxDCA and MaxBreg}
\While{$k<{\rm{MaxDCA}}$ }
{{$b_x=b_y=0$, $p=0$\;}
\While{$p<{\rm{MaxBreg}}$}{
$u=\left(\mu+\beta \nabla^{\rm{T}}\nabla\right)^{-1}\left(\mu y+\beta D_x^{\rm{T}}(d_x-b_x)+\beta D_y^{\rm{T}}(d_y-b_y)\right)$\;
$d_x=\text{shrink}\left(D_xu+b_x+\alpha D_xX^k/\beta,1/\beta \right)$\;
$d_y=\text{shrink}\left(D_yu+b_y+\alpha D_yX^k/\beta,1/\beta \right)$\;
$b_x=b_x+D_xu-d_x$\;
$b_y=b_y+D_yu-d_y$\;
$p\leftarrow p+1$\;
}
$X^{k}=u$\;
$k\leftarrow k+1$\;
}
\end{algorithm}

To reproduce Figure \ref{fig:motivation}, we test the noisy \emph{Strip} image (displayed in Figure \ref{fig:motivation}) with size $128\times 128$. The parameters are set as $\alpha = 1.2$, $\mu=0.8$, $\beta=1$, MaxDCA = 10, and MaxBreg = 1000. We contaminate the test image by adding random values onto each pixel from a normal distribution with mean 0 and standard deviation 0.6, without normalizing all pixel intensities such that they are in the range of $[0,1]$.

\section{DCA for the enhanced TV model \eqref{equ:model}}\label{sec:algorithm}

We apply the mentioned DCA in \cite{tao1997convex,tao1998dc} to solve the enhanced TV model \eqref{equ:model}. We denote by $D_xX$ and $D_yX$ the horizontal and vertical components of $\nabla X$, respectively, where $D_x$ and $D_y$ can be deemed as two operators. The DCA replaces the second component $\frac{\alpha}{2}\|\nabla X\|_2^2$ of the enhanced TV regularization term \eqref{equ:penalty} by a linear majorant $\left\langle X-X^k,\xi^k\right\rangle$,
where $\xi^k \in \partial\left(\frac{\alpha}{2}\|\nabla X\|_2^2\right)=\{\alpha \nabla^{\text{T}}\nabla X^k\}$,
and then solves the resulting convex optimization problem to generate the iterate $X^{k+1}$. Ignoring the constant term $\langle X^k,\xi^k\rangle$ in the objective function, the iterative scheme of the DCA reads as
\begin{equation}\label{equ:subproblem}
\begin{split}
X^{k+1}\in\arg\min_{X\in\mathbb{C}^{N\times N}} \left\{\|D_x X\|_1+\|D_yX\|_1-\alpha\langle D_xX,D_xX^k\rangle - \alpha\langle D_y X,D_yX^k\rangle~\text{s.t.}~\|{\mathcal{M}} X-y\|_2\leq \tau\right\}.
\end{split}\end{equation}
Convergence of the DCA \eqref{equ:subproblem} can be found in, e.g., \cite{an2021springback,tao1997convex,tao1998dc}.
Recall that a convex function $F:\mathbb{R}^d\rightarrow\mathbb{R}$ is said to be $\rho$-\emph{strongly convex} if $F(x)-\frac{\rho}{2}\|x\|_2^2$ is convex on $\mathbb{R}^d$. A simple but critical fact ensuring the convergence is that the component $\frac{\alpha}{2}\|\nabla X\|_2^2$ is strongly convex either if $X$ is mean-zero or if $X$ contains zero-valued pixels (cf. the classical Sobolev inequality \eqref{equ:classicalsobolev} and Equation \eqref{equ:22}).

To solve \eqref{equ:subproblem}, we suggest using the benchmark alternating direction method of multipliers (ADMM) in \cite{MR388811}. Clearly, $X^{k+1}$ is also a solution to the reformulated problem
\begin{equation}\label{equ:subproblem1}
\begin{split}
\min\quad&\|d_x\|_1+\|d_y\|_1-\alpha\langle d_x,D_xX^k\rangle - \alpha\langle d_y,D_yX^k\rangle,\\
\text{s.t.}\quad&{\mathcal{M}} X - y-z = 0,\\
& z\in\mathcal{B}(0,\tau):=\{x\in\mathbb{R}^m:\|x\|_2\leq\tau\},\\
& D_xX = d_x,\quad D_yX=d_y.
\end{split}\end{equation}
Introducing three Lagrange multipliers $\lambda$, $b_x$, and $b_y$, we write the augmented Lagrangian function of \eqref{equ:subproblem1} as
\begin{equation*}\begin{split}
\mathcal{L}_{\beta,\mu}(X,d_x,d_y,z,b_x,b_y,\lambda):=&\|d_x\|_1+\|d_y\|_1-\alpha\langle d_x,D_xX^k\rangle - \alpha\langle d_y,D_yX^k\rangle+\frac{\mu}{2}\|z-({\mathcal{M}} X-y)-\lambda\|_2^2\\
&+\frac{\beta}{2}\|d_x-D_xX-b_x\|_2^2+\frac{\beta}{2}\|d_y-D_yX-b_y\|_2^2,
\end{split}\end{equation*}
where $\mu,\beta>0$ are penalty parameters. Implementations of the ADMM to \eqref{equ:subproblem} are included as Algorithm \ref{alg:DCA} below, in which MaxDCA denotes the maximum number of the DCA iterations, MaxADMM is the maximum number of the ADMM iterations for \eqref{equ:subproblem1} with a given $X^k$, and ``tol" is the tolerance for the DCA iterations.
\begin{algorithm}[htbp]\label{alg:DCA}
\caption{DCA for the enhanced TV model \eqref{equ:model}}
\KwIn{Define $X^0 = 0$, $z=0$, $k=0$, $d_x=d_y=0$, MaxDCA, MaxADMM, and tol}
\While{$k<{\rm{MaxDCA}}$ and $\|X^{k}-X^{k-1}\|_2>{\rm{tol}}$ }
{{$b_x=b_y=0$, $p=0$\;}
\While{$p<{\rm{MaxADMM}}$}{
$u=\left(\mu{\mathcal{M}}^*{\mathcal{M}}+\beta \nabla^{\rm{T}}\nabla\right)^{-1}\left(\mu {\mathcal{M}}^* (y-z-\lambda)+\beta D_x^{\rm{T}}(d_x-b_x)+\beta D_y^{\rm{T}}(d_y-b_y)\right)$\;
$d_x=\text{shrink}\left(D_xu+b_x+\alpha D_xX^k/\beta,1/\beta \right)$\;
$d_y=\text{shrink}\left(D_yu+b_y+\alpha D_yX^k/\beta,1/\beta \right)$\;
$z = \mathcal{P}_{\mathcal{B}(0,\tau)}({\mathcal{M}} u-y+\lambda)$\;
$b_x=b_x+D_xu-d_x$\;
$b_y=b_y+D_yu-d_y$\;
$\lambda = \lambda+({\mathcal{M}} u-y)-z$\;
$p\leftarrow p+1$\;
}
$X^{k}=u$\;
$k\leftarrow k+1$\;
}
\end{algorithm}

In our numerical experiments, to implement Algorithm \ref{alg:DCA}, we set $\mu=10^3$, $\beta=10$, MaxDCA = 15, tol = $10^{-10}$ (for noise-free measurements) or $10^{-3}$ (for noisy measurements), and MaxADMM = 1,000. For the $\text{TV}_{\text{a}}-\text{TV}_{\text{i}}$ model in \cite{lou2015weighted}, we use the same penalty parameters and stopping criterion for running the DCA; and for the split Bregman method in solving the DCA subproblem, we set the maximum numbers of outer and inner iterations as 50 and 20, respectively. The parameters for Bregman iterations were suggested in \cite{lou2015weighted}, and they coincide with the maximum number of the inner ADMM iterations in Algorithm \ref{alg:DCA}, as $50\times 20 =1,000$. For the TV model \eqref{equ:TVmodel}, we adopt the same penalty parameters and tolerance for outer iterations. We set the maximal numbers of outer and inner iterations to be 50 and 200, respectively; both numbers were suggested in \cite{lou2015weighted}.

\bibliographystyle{siam}
\bibliography{myref}

\begin{thebibliography}{10}

\bibitem{adcock2021improved}
{\sc B.~Adcock, N.~Dexter, and Q.~Xu}, {\em Improved recovery guarantees and
  sampling strategies for {TV} minimization in compressive imaging}, SIAM
  Journal on Imaging Sciences, 14 (2021), pp.~1149--1183.

\bibitem{adcock2017breaking}
{\sc B.~Adcock, A.~C. Hansen, C.~Poon, and B.~Roman}, {\em Breaking the
  coherence barrier: A new theory for compressed sensing}, Forum of
  Mathematics, Sigma, 5 (2017).
\newblock Paper No. e4, 84 pages.

\bibitem{alvarez1994signal}
{\sc L.~Alvarez and L.~Mazorra}, {\em Signal and image restoration using shock
  filters and anisotropic diffusion}, SIAM Journal on Numerical Analysis, 31
  (1994), pp.~590--605.

\bibitem{an2021springback}
{\sc C.~An, H.-N. Wu, and X.~Yuan}, {\em The springback penalty for robust
  signal recovery}, Applied and Computational Harmonic Analysis, 61 (2022),
  pp.~319--346.

\bibitem{benning2013higher}
{\sc M.~Benning, C.~Brune, M.~Burger, and J.~M{\"u}ller}, {\em Higher-order
  {TV} methods---enhancement via {B}regman iteration}, Journal of Scientific
  Computing, 54 (2013), pp.~269--310.

\bibitem{blomgren1997total}
{\sc P.~Blomgren, T.~F. Chan, P.~Mulet, and C.-K. Wong}, {\em Total variation
  image restoration: Numerical methods and extensions}, in Proceedings of
  International Conference on Image Processing, IEEE, 1997, pp.~384--387.

\bibitem{bredies2010total}
{\sc K.~Bredies, K.~Kunisch, and T.~Pock}, {\em Total generalized variation},
  SIAM Journal on Imaging Sciences, 3 (2010), pp.~492--526.

\bibitem{cai2015guarantees}
{\sc J.-F. Cai and W.~Xu}, {\em Guarantees of total variation minimization for
  signal recovery}, Information and Inference: A Journal of the IMA, 4 (2015),
  pp.~328--353.

\bibitem{candes2006robust}
{\sc E.~J. Cand{\`e}s, J.~Romberg, and T.~Tao}, {\em Robust uncertainty
  principles: Exact signal reconstruction from highly incomplete frequency
  information}, IEEE Transactions on Information Theory, 52 (2006),
  pp.~489--509.

\bibitem{candes2005decoding}
{\sc E.~J. Cand{\`e}s and T.~Tao}, {\em Decoding by linear programming}, IEEE
  Transactions on Information Theory, 51 (2005), pp.~4203--4215.

\bibitem{candes2006near}
\leavevmode\vrule height 2pt depth -1.6pt width 23pt, {\em Near-optimal signal
  recovery from random projections: {U}niversal encoding strategies?}, IEEE
  Transactions on Information Theory, 52 (2006), pp.~5406--5425.

\bibitem{chambolle2004algorithm}
{\sc A.~Chambolle}, {\em An algorithm for total variation minimization and
  applications}, Journal of Mathematical Imaging and Vision, 20 (2004),
  pp.~89--97.

\bibitem{chambolle2005total}
\leavevmode\vrule height 2pt depth -1.6pt width 23pt, {\em Total variation
  minimization and a class of binary {MRF} models}, in International Workshop
  on Energy Minimization Methods in Computer Vision and Pattern Recognition,
  Springer, Berlin, Heidelberg, 2005, pp.~136--152.

\bibitem{MR2731599}
{\sc A.~Chambolle, V.~Caselles, D.~Cremers, M.~Novaga, and T.~Pock}, {\em An
  introduction to total variation for image analysis}, in Theoretical
  Foundations and Numerical Methods for Sparse Recovery, De Gruyter, Berlin,
  2010, pp.~263--340.

\bibitem{chambolle1997image}
{\sc A.~Chambolle and P.-L. Lions}, {\em Image recovery via total variation
  minimization and related problems}, Numerische Mathematik, 76 (1997),
  pp.~167--188.

\bibitem{chambolle2016introduction}
{\sc A.~Chambolle and T.~Pock}, {\em An introduction to continuous optimization
  for imaging}, Acta Numerica, 25 (2016), pp.~161--319.

\bibitem{chambolle2021approximating}
\leavevmode\vrule height 2pt depth -1.6pt width 23pt, {\em Approximating the
  total variation with finite differences or finite elements}, in Handbook of
  Numerical Analysis, vol.~22, Elsevier, 2021, pp.~383--417.

\bibitem{chambolle2021learning}
\leavevmode\vrule height 2pt depth -1.6pt width 23pt, {\em Learning consistent
  discretizations of the total variation}, SIAM Journal on Imaging Sciences, 14
  (2021), pp.~778--813.

\bibitem{chan2005aspects}
{\sc T.~F. Chan and S.~Esedoḡlu}, {\em Aspects of total variation regularized
  ${L}^1$ function approximation}, SIAM Journal on Applied Mathematics, 65
  (2005), pp.~1817--1837.

\bibitem{chan2000high}
{\sc T.~F. Chan, A.~Marquina, and P.~Mulet}, {\em High-order total
  variation-based image restoration}, SIAM Journal on Scientific Computing, 22
  (2000), pp.~503--516.

\bibitem{chartrand2007exact}
{\sc R.~Chartrand}, {\em Exact reconstruction of sparse signals via nonconvex
  minimization}, IEEE Signal Processing Letters, 14 (2007), pp.~707--710.

\bibitem{MR1411676}
{\sc J.~H. Conway and R.~K. Guy}, {\em The Book of Numbers}, Copernicus, New
  York, 1996.

\bibitem{donoho2006compressed}
{\sc D.~L. Donoho}, {\em Compressed sensing}, IEEE Transactions on Information
  Theory, 52 (2006), pp.~1289--1306.

\bibitem{esedoglu2004decomposition}
{\sc S.~Esedoḡlu and S.~Osher}, {\em Decomposition of images by the
  anisotropic {R}udin--{O}sher--{F}atemi model}, Communications on Pure and
  Applied Mathematics, 57 (2004), pp.~1609--1626.

\bibitem{fan2001variable}
{\sc J.~Fan and R.~Li}, {\em Variable selection via nonconcave penalized
  likelihood and its oracle properties}, Journal of the American Statistical
  Association, 96 (2001), pp.~1348--1360.

\bibitem{fannjiang2010compressed}
{\sc A.~C. Fannjiang, T.~Strohmer, and P.~Yan}, {\em Compressed remote sensing
  of sparse objects}, SIAM Journal on Imaging Sciences, 3 (2010), pp.~595--618.

\bibitem{foucart2009sparsest}
{\sc S.~Foucart and M.-J. Lai}, {\em Sparsest solutions of underdetermined
  linear systems via $\ell_q$-minimization for $0<q\leq 1$}, Applied and
  Computational Harmonic Analysis, 26 (2009), pp.~395--407.

\bibitem{galdran2015enhanced}
{\sc A.~Galdran, J.~Vazquez-Corral, D.~Pardo, and M.~Bertalmio}, {\em Enhanced
  variational image dehazing}, SIAM Journal on Imaging Sciences, 8 (2015),
  pp.~1519--1546.

\bibitem{gilboa2002forward}
{\sc G.~Gilboa, N.~Sochen, and Y.~Y. Zeevi}, {\em Forward-and-backward
  diffusion processes for adaptive image enhancement and denoising}, IEEE
  Transactions on Image Processing, 11 (2002), pp.~689--703.

\bibitem{MR388811}
{\sc R.~Glowinski and A.~Marrocco}, {\em Sur l'approximation, par
  \'{e}l\'{e}ments finis d'ordre un, et la r\'{e}solution, par
  p\'{e}nalisation-dualit\'{e}, d'une classe de probl\`emes de {D}irichlet non
  lin\'{e}aires}, Rev. Fran\c{c}aise Automat. Informat. Recherche
  Op\'{e}rationnelle S\'{e}r. Rouge Anal. Num\'{e}r., 9 (1975), pp.~41--76.

\bibitem{goldstein2009split}
{\sc T.~Goldstein and S.~Osher}, {\em The split {B}regman method for
  {L}1-regularized problems}, SIAM Journal on Imaging Sciences, 2 (2009),
  pp.~323--343.

\bibitem{krahmer2017total}
{\sc F.~Krahmer, C.~Kruschel, and M.~Sandbichler}, {\em Total variation
  minimization in compressed sensing}, in Compressed Sensing and its
  Applications, Birkh\"{a}user/Springer, Cham, 2017, pp.~333--358.

\bibitem{krahmer2011new}
{\sc F.~Krahmer and R.~Ward}, {\em New and improved {J}ohnson--{L}indenstrauss
  embeddings via the restricted isometry property}, SIAM Journal on
  Mathematical Analysis, 43 (2011), pp.~1269--1281.

\bibitem{krahmer2013stable}
\leavevmode\vrule height 2pt depth -1.6pt width 23pt, {\em Stable and robust
  sampling strategies for compressive imaging}, IEEE Transactions on Image
  Processing, 23 (2014), pp.~612--622.

\bibitem{li2020ell1}
{\sc P.~Li, W.~Chen, H.~Ge, and M.~K.-P. Ng}, {\em $\ell_1-\alpha\ell_2$
  minimization methods for signal and image reconstruction with impulsive noise
  removal}, Inverse Problems, 36 (2020), p.~055009.

\bibitem{lou2015weighted}
{\sc Y.~Lou, T.~Zeng, S.~Osher, and J.~Xin}, {\em A weighted difference of
  anisotropic and isotropic total variation model for image processing}, SIAM
  Journal on Imaging Sciences, 8 (2015), pp.~1798--1823.

\bibitem{lustig2007sparse}
{\sc M.~Lustig, D.~Donoho, and J.~M. Pauly}, {\em Sparse {MRI}: {T}he
  application of compressed sensing for rapid {MR} imaging}, Magnetic Resonance
  in Medicine, 58 (2007), pp.~1182--1195.

\bibitem{lustig2008compressed}
{\sc M.~Lustig, D.~L. Donoho, J.~M. Santos, and J.~M. Pauly}, {\em Compressed
  sensing {MRI}}, IEEE Signal Processing Magazine, 25 (2008), pp.~72--82.

\bibitem{mendelson2007reconstruction}
{\sc S.~Mendelson, A.~Pajor, and N.~Tomczak-Jaegermann}, {\em Reconstruction
  and subgaussian operators in asymptotic geometric analysis}, Geometric and
  Functional Analysis, 17 (2007), pp.~1248--1282.

\bibitem{MR2139257}
{\sc J.~S. Moll}, {\em The anisotropic total variation flow}, Mathematische
  Annalen, 332 (2005), pp.~177--218.

\bibitem{mollenhoff2015primal}
{\sc T.~M\"{o}llenhoff, E.~Strekalovskiy, M.~Moeller, and D.~Cremers}, {\em The
  primal-dual hybrid gradient method for semiconvex splittings}, SIAM Journal
  on Imaging Sciences, 8 (2015), pp.~827--857.

\bibitem{needell2013near}
{\sc D.~Needell and R.~Ward}, {\em Near-optimal compressed sensing guarantees
  for total variation minimization}, IEEE Transactions on Image Processing, 22
  (2013), pp.~3941--3949.

\bibitem{needell2013stable}
\leavevmode\vrule height 2pt depth -1.6pt width 23pt, {\em Stable image
  reconstruction using total variation minimization}, SIAM Journal on Imaging
  Sciences, 6 (2013), pp.~1035--1058.

\bibitem{nikolova2002minimizers}
{\sc M.~Nikolova}, {\em Minimizers of cost-functions involving nonsmooth
  data-fidelity terms. application to the processing of outliers}, SIAM Journal
  on Numerical Analysis, 40 (2002), pp.~965--994.

\bibitem{MR3560068}
\leavevmode\vrule height 2pt depth -1.6pt width 23pt, {\em Energy minimization
  methods}, in Handbook of Mathematical Methods in Imaging, Springer, New York,
  2015, pp.~157--204.

\bibitem{osher1990feature}
{\sc S.~Osher and L.~I. Rudin}, {\em Feature-oriented image enhancement using
  shock filters}, SIAM Journal on Numerical Analysis, 27 (1990), pp.~919--940.

\bibitem{pierre2017variational}
{\sc F.~Pierre, J.-F. Aujol, A.~Bugeau, G.~Steidl, and V.-T. Ta}, {\em
  Variational contrast enhancement of gray-scale and rgb images}, Journal of
  Mathematical Imaging and Vision, 57 (2017), pp.~99--116.

\bibitem{poon2015role}
{\sc C.~Poon}, {\em On the role of total variation in compressed sensing}, SIAM
  Journal on Imaging Sciences, 8 (2015), pp.~682--720.

\bibitem{rauhut2012restricted}
{\sc H.~Rauhut, J.~Romberg, and J.~A. Tropp}, {\em Restricted isometries for
  partial random circulant matrices}, Applied and Computational Harmonic
  Analysis, 32 (2012), pp.~242--254.

\bibitem{rauhut2012sparse}
{\sc H.~Rauhut and R.~Ward}, {\em Sparse {L}egendre expansions via
  $\ell_1$-minimization}, Journal of Approximation Theory, 164 (2012),
  pp.~517--533.

\bibitem{rudelson2008sparse}
{\sc M.~Rudelson and R.~Vershynin}, {\em On sparse reconstruction from
  {F}ourier and {G}aussian measurements}, Communications on Pure and Applied
  Mathematics, 61 (2008), pp.~1025--1045.

\bibitem{rudin1992nonlinear}
{\sc L.~I. Rudin, S.~Osher, and E.~Fatemi}, {\em Nonlinear total variation
  based noise removal algorithms}, Physica D: Nonlinear Phenomena, 60 (1992),
  pp.~259--268.

\bibitem{MR2544023}
{\sc S.~Setzer and G.~Steidl}, {\em Variational methods with higher-order
  derivatives in image processing}, in Approximation theory {XII}: {S}an
  {A}ntonio 2007, Mod. Methods Math., Nashboro Press, Brentwood, TN, 2008,
  pp.~360--385.

\bibitem{setzer2011infimal}
{\sc S.~Setzer, G.~Steidl, and T.~Teuber}, {\em Infimal convolution
  regularizations with discrete $\ell_1$-type functionals}, Communications in
  Mathematical Sciences, 9 (2011), pp.~797--827.

\bibitem{strong2003edge}
{\sc D.~Strong and T.~F. Chan}, {\em Edge-preserving and scale-dependent
  properties of total variation regularization}, Inverse Problems, 19 (2003),
  p.~S165.

\bibitem{tao1997convex}
{\sc P.~D. Tao and L.~T.~H. An}, {\em Convex analysis approach to {DC}
  programming: theory, algorithms and applications}, Acta Mathematica
  Vietnamica, 22 (1997), pp.~289--355.

\bibitem{tao1998dc}
\leavevmode\vrule height 2pt depth -1.6pt width 23pt, {\em A {DC} optimization
  algorithm for solving the trust-region subproblem}, SIAM Journal on
  Optimization, 8 (1998), pp.~476--505.

\bibitem{MR0455365}
{\sc A.~N. Tikhonov and V.~Y. Arsenin}, {\em Solutions of ill-posed problems},
  John Wiley \& Sons; Washington, D.C, 1977.
\newblock Translated from the Russian, Preface by translation editor Fritz
  John.

\bibitem{welk2009theoretical}
{\sc M.~Welk, G.~Gilboa, and J.~Weickert}, {\em Theoretical foundations for
  discrete forward-and-backward diffusion filtering}, in International
  Conference on Scale Space and Variational Methods in Computer Vision,
  Springer, 2009, pp.~527--538.

\bibitem{welk2008locally}
{\sc M.~Welk, G.~Steidl, and J.~Weickert}, {\em Locally analytic schemes: {A}
  link between diffusion filtering and wavelet shrinkage}, Applied and
  Computational Harmonic Analysis, 24 (2008), pp.~195--224.

\bibitem{welk2005pde}
{\sc M.~Welk, D.~Theis, T.~Brox, and J.~Weickert}, {\em {PDE}-based
  deconvolution with forward-backward diffusivities and diffusion tensors}, in
  International Conference on Scale-Space Theories in Computer Vision,
  Springer, 2005, pp.~585--597.

\bibitem{welk2007theoretical}
{\sc M.~Welk, J.~Weickert, and I.~Gali{\'c}}, {\em Theoretical foundations for
  spatially discrete 1-{D} shock filtering}, Image and Vision Computing, 25
  (2007), pp.~455--463.

\bibitem{welk2018discrete}
{\sc M.~Welk, J.~Weickert, and G.~Gilboa}, {\em A discrete theory and efficient
  algorithms for forward-and-backward diffusion filtering}, Journal of
  Mathematical Imaging and Vision, 60 (2018), pp.~1399--1426.

\bibitem{yin2015minimization}
{\sc P.~Yin, Y.~Lou, Q.~He, and J.~Xin}, {\em Minimization of $\ell_{1-2}$ for
  compressed sensing}, SIAM Journal on Scientific Computing, 37 (2015),
  pp.~A536--A563.

\bibitem{zhang2018minimization}
{\sc S.~Zhang and J.~Xin}, {\em Minimization of transformed $l_1$ penalty:
  theory, difference of convex function algorithm, and robust application in
  compressed sensing}, Mathematical Programming, 169 (2018), pp.~307--336.

\end{thebibliography}
\clearpage

\end{document}